\documentclass{aa}
\usepackage{txfonts}
\usepackage{natbib}
\bibpunct{(}{)}{;}{a}{}{,}
\usepackage{graphicx}
\usepackage{placeins}

\newcommand{\teff}{$T_{\rm{eff}}$}

\newcommand{\lL}{\ifmmode \log \frac{L}{L_{\sun}} \else $\log \frac{L}{L_{\sun}}$\fi}
\newcommand{\mdot}{$\dot{M}$}

\newcommand{\vinf}{$\varv_{\infty}$}

\newcommand{\vesc}{$\varv_{\rm esc}$}
\newcommand{\kms}{km~s$^{-1}$}
\newcommand{\msun}{M$_{\sun}$}
\newcommand{\zsun}{Z$_{\sun}$}
\newcommand{\lya}{Ly$\alpha$}
\newcommand{\heiiuv}{\ion{He}{ii}~1640}
\newcommand{\heiiopt}{\ion{He}{ii}~4686}
\newcommand{\civuv}{\ion{C}{iv}~1550}
\newcommand{\nivuv}{\ion{N}{iv}~1720}

\begin{document}

\title{Spectroscopic evolution of very massive stars at Z = 1/2.5 \zsun}
\author{F. Martins\inst{1}
\and A. Palacios\inst{1}
}
\institute{LUPM, Universit\'e de Montpellier, CNRS, Place Eug\`ene Bataillon, F-34095 Montpellier, France  \\
}

\offprints{prenom nom\\ \email{fabrice.martins@umontpellier.fr}}

\date{Received / Accepted }

\abstract
{Stars with masses in excess of 100~\msun\ are observed in the Local Universe, but they remain rare objects. Because of the shape of the mass function, they are expected to be present only in the most massive and youngest clusters. They may thus be formed in number in highly star-forming galaxies.}
{Very massive stars (VMSs) experience strong stellar winds that are stronger than those of their less massive OB-type counterparts. These strong winds therefore need to be taken into account in evolutionary models and synthetic spectra to properly predict the appearance of VMS. }
{We present evolutionary models computed with the code \emph{STAREVOL.}  They include a recent mass-loss recipe that is relevant for VMSs. We subsequently calculated atmosphere models and synthetic spectra along the resulting tracks with the code \emph{CMFGEN}. We studied stars with masses between 150 and 400~\msun\ and focused on a metallicity Z~=~1/2.5~\zsun. We studied the impact of our VMS spectra on the spectral energy distribution of young starbursts.}
{We show that the optical and UV range is dominated by \heiiopt\ and \heiiuv\ emission for almost the entire main-sequence evolution of VMSs, in contrast to less massive stars. In the UV spectral range, carbon, nitrogen, and iron lines shape the spectra of VMSs, which appear for most of their evolution as WNh objects. The morphology of the synthetic spectra is similar to that of VMSs in the Large Magellanic Cloud. We show that stars with masses higher than 100~\msun\ emit nearly as much light as all other stars in young starbursts. The integrated UV spectrum of these starbursts is significantly affected by the presence of VMSs. }
{We conclude that a consistent treatment of the evolution and the atmospheres of VMSs is mandatory to properly study spatially unresolved regions of intense star formation. We make our synthetic spectra and spectral energy distributions available to the community.}


\authorrunning{Martins \& Palacios}
\titlerunning{Very massive stars at Z = 1/2.5 \zsun}

\maketitle

\section{Introduction}
\label{s_intro}

Very massive stars (VMSs) have masses in excess of 100~\msun\ \citep{vink15vms}. It is an open question how far above this threshold the mass of stars extends. Analyses of the stellar content of young massive clusters in the Galaxy and the Magellanic Clouds have lead to the prediction of an upper mass cut-off of 150-200~\msun\ \citep{weidner04,oey05,figer05}. Dynamical measurements of stellar masses in binary systems are so far consistent with this limit: none of the most massive stars observed in spectroscopic binaries has a mass higher than 150~\msun\ \citep{schnurr08,tehrani19,shenar21}. The current record holder for an eclipsing system, in which the dynamical mass can be probed directly, without an assumption about the inclination,  is NGC3603-A1, in which the primary component has a present-day mass of 116$\pm$31~\msun\ \citep{schnurr08}. Masses can also be obtained from stellar luminosity and comparison to theoretical predictions of evolutionary models. This can be done for binary components and for single stars \citep{martins15vms}. Using this method, \citet{crowther10} reported masses up to 320~\msun\ in the Large Magellanic Cloud (LMC) cluster R136. Although these mass estimates have been revised since then, they are still higher than 200~\msun\ for a few objects \citep{besten20}.

Since VMSs evolve rapidly and because of the shape of the initial mass function (IMF), young massive star clusters are the preferred places to search for VMSs. \citet{wofford14} presented HST/COS ultraviolet (UV) spectroscopy of these clusters in the blue compact dwarf galaxy NGC~3125, at a metallicity close to that of the LMC. Standard population synthesis models with a Salpeter IMF extending up to 100~\msun\ were unable to reproduce all lines, in particular the broad \heiiuv\ emission. Because the spectrum of cluster NGC3215-A1 is similar to that of R136 in the LMC, where VMSs are observed, the authors concluded that stars with masses in excess of 100~\msun\ are a necessary ingredient of population synthesis models \citep[see also][]{smith16,senchyna17,leitherer18}. Incorporating updated evolutionary tracks of VMSs in these models, \citet{senchyna21} confirmed that better agreement between observed and predicted equivalent width (EW) of UV lines is obtained for extreme star-forming regions. However, they relied on atmosphere models that are not specifically designed for VMSs. 
On even broader spatial scales, \citet{gos21} studied the [O/Fe] ratio of Galactic thin- and thick-disk stars and concluded that very massive stars up to 350~\msun\ are required to explain the properties of the latter group of stars.
It is thus necessary to develop tailored evolutionary and atmosphere models of VMSs to study young stellar populations in the Local Universe.

A key ingredient of modelling VMSs is their strong stellar winds. \citet{martins08} showed that the mass-loss rates of the most massive stars in the Galactic Arches cluster scaled with luminosity, albeit with a significant offset compared to the relation followed by O-type stars \citep[see also][]{hainich14}. In parallel, \citet{gh08} investigated the driving of stellar winds in late-type WN stars. They concluded that very luminous, H-rich WNh stars developed optically thick radiatively driven winds mainly because of their proximity to the Eddington limit. Subsequent investigations confirmed that VMSs have higher mass-loss rates than more classical and less massive O-type stars \citep{vink11,graef11}. Using observations of young massive stars in 30~Dor, \citet{besten14} derived a mass-loss rate recipe for VMSs in the LMC \citep[see also][]{besten20a}. \citet{graef21} used a combination of this recipe and that of \citet{vink01} to study the impact of mass loss on the evolution of VMSs in the LMC. However, the spectral appearance was not studied. Conversely, \citet{gv15} presented the first synthetic spectra of VMSs at different metallicities, but  provided only snapshot spectra for a specific set of stellar parameters (only one value of effective temperature and luminosity).

In the present paper we describe a pilot study aiming at providing synthetic spectra of VMSs along their evolutionary paths. We concentrate on an LMC-like metallicity to benefit from the mass-loss recipe of \citet{besten14} and \citet{graef21}. We first describe our evolutionary models and how they compare to previous studies (Sect.~\ref{s_ev}). After this, we present the calculation of synthetic spectra and describe the spectroscopic evolutionary sequences in the optical and UV spectral ranges (Sect.~\ref{s_spectro}). We explore the impact of our models on population synthesis in Sect.~\ref{s_popsyn}, and we conclude in Sect.~\ref{s_conc}.

\section{Evolutionary models}
\label{s_ev}

In this section we present the main ingredients we used to calculate evolutionary tracks of VMSs. We describe in detail our implementation of the mass-loss recipe of \citet{graef21} and then compare our results to grids of models that are available in the literature.

\subsection{Physical ingredients}
\label{s_setup}

As in \cite{mp21}, we computed non-rotating evolutionary models with the code \emph{STAREVOL} \citep{siess00,lagarde12,amard19}. We modelled stars with initial masses of 150, 200, 250, 300, and 400~\msun. We recall here the main physical ingredients relevant for this study. We assumed an Eddington grey atmosphere as outer boundary condition to the stellar structure equations. We used the
\cite{agss09} solar chemical composition as a reference, with Z$_\odot$~=~0.0134. A calibration of the solar model with these input physics leads to an initial helium mass fraction Y~=~0.2687 at solar metallicity. We used the corresponding constant slope $\Delta$Y/$\Delta$Z~=~1.60 (with the primordial abundance Y$_0$= 0.2463 based on WMAP-SBBN by \citealt{coc2004}) to compute the initial helium mass fraction at Z~=~$5.50 \times 10^{-3}$ , corresponding to [Fe/H]~=~-0.4 and typical of the LMC stars. We then scaled all the abundances accordingly. The OPAL opacities used for these models comply to this scaled distribution of nuclides. We did not include specific $\alpha$-element enhancement in our models. We described the convective instability using the mixing-length theory with $\alpha_{\rm MLT}$~=~1.6304, and we used the Schwarzschild instability criterion to
define the boundaries of convective regions. We added a step
overshoot at the convective core edge and adopted $\alpha_{\rm ov} = 0.1 {\rm H_p}$, with H$_{\rm p}$ being the pressure scale height. We used the thermonuclear reaction rates from the NACRE II compilation \citep{nacre2} for mass number A < 16, and those from the NACRE compilation \citep{nacre} for more massive nuclei up to Ne. The proton captures on nuclei more massive than Ne are from \cite{longland2010} or \cite{iliadis2001}. The network was generated via the NetGen server \citep{xu13}. 

\subsection{Mass-loss treatment}\label{sec:massloss}
\label{s_mdot}

 We implemented the mass-loss treatment prescribed by \cite{graef21} in our models. The mass-loss recipe of VMSs depends on the optical thickness of their winds. While they appear as O stars, their winds are optically thin. However, when they enter the WNh phase (see Sect.~\ref{s_opt} for a definition) while still on the main sequence, their winds become optically thick and are not properly described by the mass loss of classical core He-burning Wolf-Rayet stars. \\\cite{graef21} thus proposed a criterion to switch between the optically thin wind regime (O-type star) and the optically thick wind regime (WNh-type star). This criterion is tailored for LMC VMSs; it was fitted on observed mass-loss rates for VMSs in 30~Dor as derived by \cite{besten14}. It is based on the sonic-point optical depth $\tau_s$ in the wind, which for a stellar model with surface mass $M$, luminosity $L$, radius $R$, and hydrogen mass fraction $X$ writes 
 \begin{equation}
 \tau_s \approx \frac{\dot{M} v_\infty}{L/c} \left(1 + \frac{v^2_{\rm esc}}{v^2_\infty}\right)
 \label{eq:sonicpoint}
 ,\end{equation}
 where $v_\infty$ and $v_{\rm esc}$ are the terminal and escape velocity in the wind, respectively, and $v_\infty/v^{\rm eff}_{\rm esc}$ = 2.51 $\pm$ 0.27  following the empirical relation for O stars with $T_{\rm eff} > 25 000$ K from \cite{lamers95}. The escape velocity $v_{\rm esc}$ is assumed to coincide with the effective escape velocity,
  \begin{equation}
     v_{\rm esc} \equiv v^{\rm eff}_{\rm esc} = \sqrt{2 G M (1-\Gamma_e)/R_{\rm eff}}
 ,\end{equation} 
  \noindent where $R_{\rm eff}$ is the effective radius such as $L = 4\pi R^2_{\rm eff}\sigma T^4_{\rm eff}$, and $\Gamma_e$ is the classical Eddington factor expressed in the case of fully ionized plasma, when the electron scattering opacity $\chi_e$ only depends on the hydrogen mass fraction $X,$ 
  
  \begin{equation}
     \Gamma_e = \frac{\chi_e L}{4 \pi c G M} = 10^{-4.813}  (1+X) \frac{L}{L_\odot}\frac{M_\odot}{M}.
 \end{equation}
 
  For models at the LMC metallicity, the criterion to decide between the two wind regimes is the following:
  
\begin{eqnarray*}
~~~\text{if} ~\tau_s < \frac{2}{3} &\text{, then} &\text{the optically thin regime} \rightarrow  \dot{M} \equiv \dot{M}_{\rm thin}\\
~~~\text{if}~ \tau_s > 1 &\text{, then the} &\text{optically thick regime} \rightarrow \dot{M} \equiv \dot{M}_{\rm thick}
\end{eqnarray*}
 In the optically thin regime, which is characteristic of the OB stars, the mass-loss rate is the rate derived by \cite{vink01} on the hot side of the bistability jump, that is, for $27 500 \leq T_{\rm eff} \leq 50 000$ K, 
  
  \begin{eqnarray} \label{eq:Mthin}
 \log(\dot{M}_{\rm thin}) = & -& 6.697 ~(\pm 0.061)\\ \nonumber
 &+& 2.194 ~(\pm 0.021) \times \log\left(\frac{L_*}{10^5}\right)\\\nonumber
&-&1.313 ~(\pm 0.046) \times  \log\left(\frac{M_*}{30}\right)\\\nonumber
&-&1.226 ~(\pm 0.037) \times \log\left(\frac{v_\infty/v_{esc}}{2}\right)\\\nonumber
&+&0 .933 ~(\pm 0.064) \times \log\left(\frac{T_{\rm eff}}{40 000}\right)\\\nonumber
&-&10.92 ~(\pm 0.90) \times {\log\left(\frac{T_{\rm eff}}{40 000}\right)}^2\\\nonumber
&+&0.85 ~(\pm 0.10) \times \log\left(\frac{Z}{Z_\odot}\right)
 \end{eqnarray}
 
\noindent where $\dot{M}$ is in $M_\odot$/yr, $L_*$ and $M_*$ are the luminosity and mass of the star in solar units, $T_{\rm eff}$ is the effective temperature in Kelvin, the ratio between terminal and escape velocities $v_\infty /v_{\rm esc}$ is taken equal to 2.6 as in Eq. (24) of \cite{vink01} (which is compatible within the error bars with the value of 2.51 used in Eq.~\ref{eq:sonicpoint}), and Z$_\odot = 0.019$. \\ 

This expression differs from that adopted by \cite{graef21}, who advocated two modifications: a global decrease in mass loss by 20 \%, and the use of a shallower slope of the mass loss as a function of L, with the replacement of the factor 2.194 in Eq.~(\ref{eq:Mthin}) by a factor 1.45. We decided not to apply the first modification because it was added in order to reproduce observed LMC trends, which is not our main goal here. We conserved the original factor by \cite{vink01} in the second modification because the modified factor of 1.45 proposed by \cite{graef21} comes from the work of \cite{besten14}, where it characterized the wind-momentum\footnote{The modified wind momentum is defined by $\dot{\rm M} v_\infty \sqrt{R_\star/R_\odot}$ following \cite{kudritz99}.} luminosity relation (see their Eq. 7 and Table 5) and not the mass-loss luminosity relation. In practice, only the earliest phase (above 53000 K) of the 150~\msun\ track is in the optically thin regime: the remainder of the 150~\msun\ track as well as the entire track for masses above 200~\msun\ are spent in the optically thick regime. Hence the choice of the mass-loss recipe in the optically thin regime has little effect on our results. \\

In the optically thick regime, the mass-loss rate comes from \cite{besten14},
\begin{equation}
\label{eq_mdot}
    \log(\dot{M}_{\rm thick}) = 5.22 \times \log(\Gamma_{\rm eff}) - 0.5 \times \log(D) -2.6
,\end{equation}

\noindent with $D = 10 $ the wind clumping factor, and $\Gamma_{\rm eff}$ is the effective Eddington factor for a rotating star taken at an intermediate latitude,
\begin{equation}
    \Gamma_{\rm eff} = \Gamma_e + 0.5 \times \Gamma_{\rm rot} = \Gamma_e + \frac{\Omega^2 R^3}{2 G M},
\end{equation}
where $\Omega$ is the surface angular velocity.
We focus on non-rotating models here so that $\Gamma_{\rm eff} \equiv \Gamma_e$.
Finally, in the intermediate regime, when $2/3 \leq \tau_s \leq 1$, we assumed a smooth transition and determined the mass-loss rate by performing a linear interpolation between $\dot{M}_{\rm thin}$ and $\dot{M}_{\rm thick}$.

\begin{figure}[t]
\centering
\includegraphics[width=0.49\textwidth]{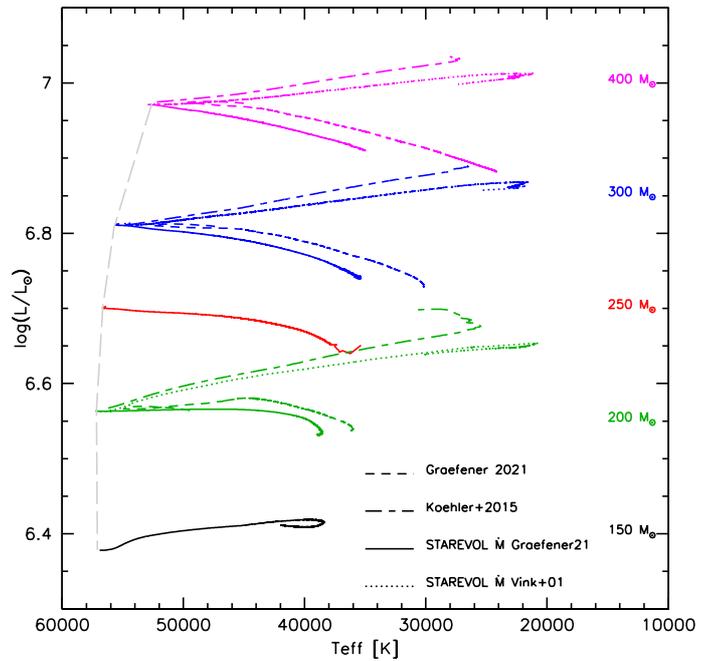}
\caption{Hertzsprung-Russell diagram (HRD) for our models with two different mass-loss treatments compared to those published in \cite{kohler15} and \cite{graef21}. See text for more details. Only the main-sequence part of the tracks up to $X_c \approx 0.01$ is shown for all models. The long dashed gray line represents the ZAMS of the \emph{STAREVOL}  models.}
\label{fig:evolcomp}
\end{figure}

\subsection{Comparison to published grids and surface chemical evolution}
\label{s_compev}
In Fig.~\ref{fig:evolcomp} we present the evolutionary paths of the non-rotating {\em STAREVOL} models as described in the sections above. We also show three models, for 200 $M_\odot$, 300 $M_\odot$ , and 400 $M_\odot$ computed with a classical mass-loss recipe\footnote{This mass-loss treatment is similar to that referred to as the Dutch scheme in the MESA code.} combining \cite{vink01} and \cite{nl00} when the models enter a Wolf-Rayet-like phase that we used in \cite{mp17}. In these latter models, we did not include wind clumping in order to be consistent with the models computed by \cite{kohler15} with the BEC code \citep{heger00,brott11}. We show the latter models in Fig.~\ref{fig:evolcomp} for comparison. Finally, we overplot the 200 $M_\odot$, 300 $M_\odot$ , and 400 $M_\odot$ non-rotating models computed by \cite{graef21} with the MESA code \citep[][and references therein]{paxton19}. In order to alleviate the figure, we limited the tracks to parts in which the central hydrogen mass fractions are larger than about 0.01. 

The tracks that include the classical mass-loss recipe shown here as dotted lines can be directly compared to those of \cite{kohler15} shown as dotted lines. The agreement is fairly good considering the different assumptions made in this study for the input physics. The  \emph{STAREVOL}  models exhibit a redder extension of the main sequence than the BEC models, but this turning point is very sensitive to both mass loss and convection treatment, so that the observed differences are not a concern. 

\begin{figure*}[ht]
\centering
\includegraphics[width=0.49\textwidth]{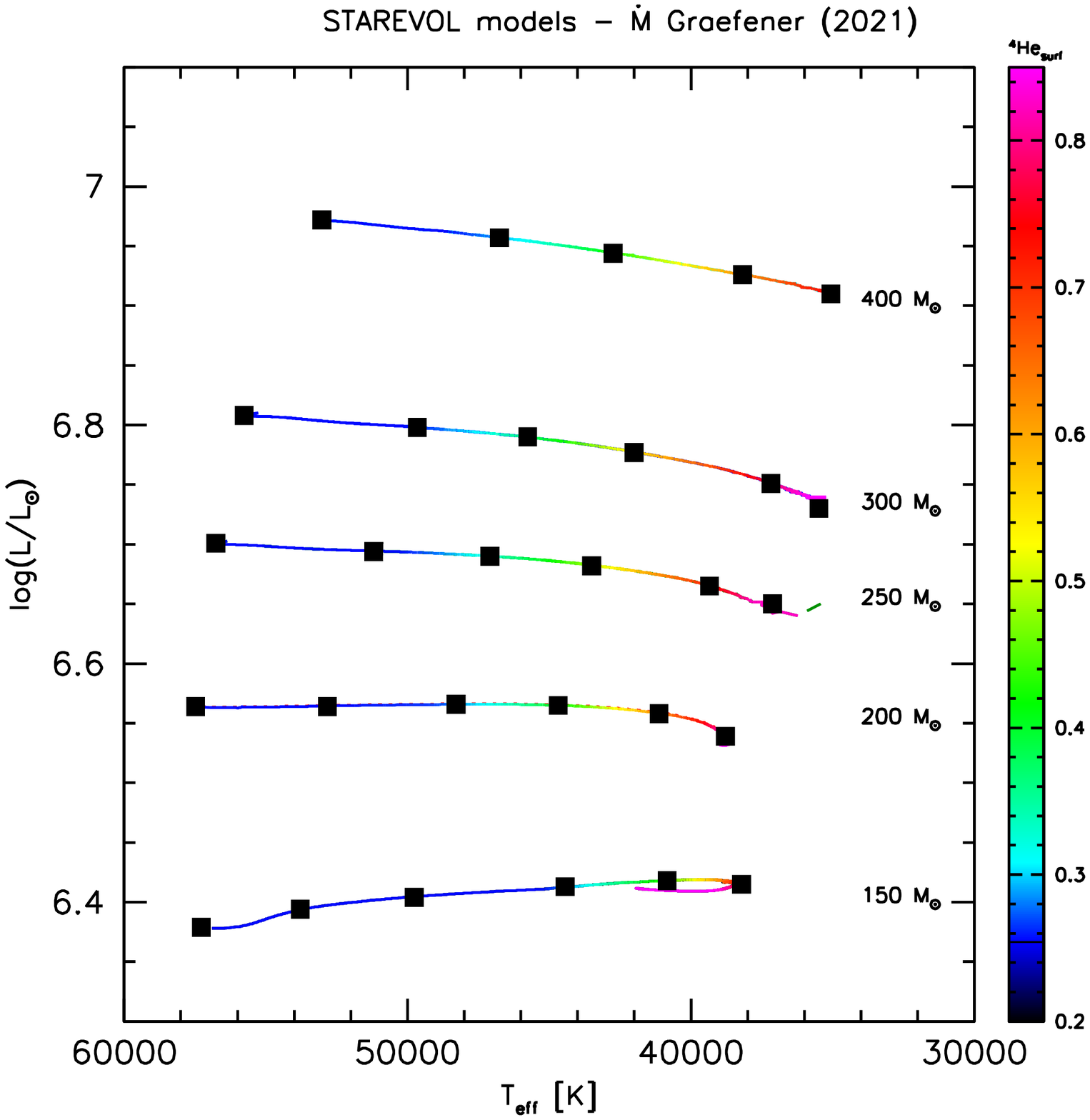}%
\includegraphics[width=0.49\textwidth]{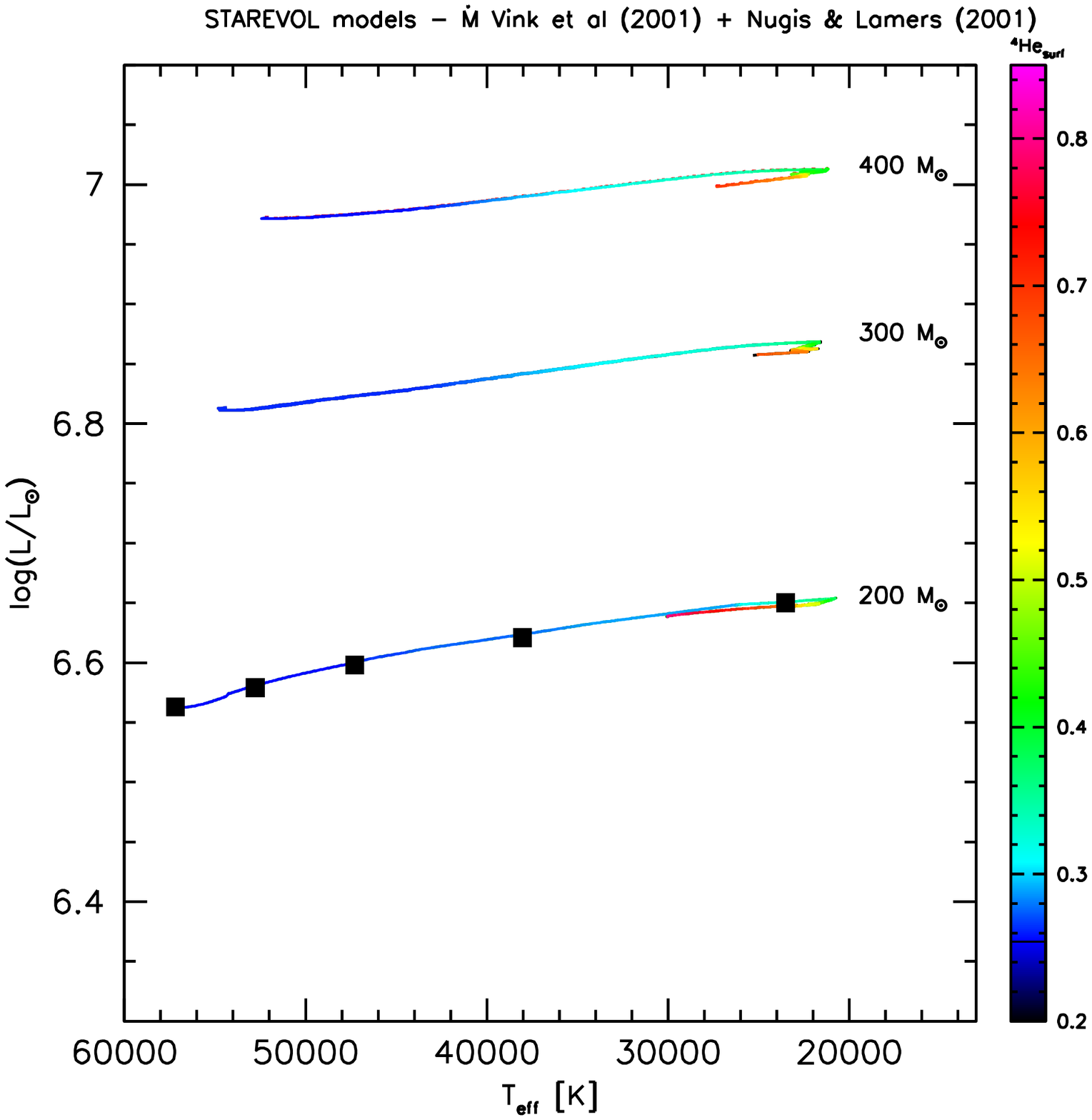}
\caption{HRD for the \emph{STAREVOL}  models on the main sequence. The most evolved tracks (150 and 200 $M_\odot$) are shown up to a central hydrogen mass fraction equal to 0.015. The evolution of the surface helium mass fraction is colour-coded on the tracks. The left panel shows the tracks obtained using the new mass-loss recipe by \cite{graef21}, and the right panel shows those using the old mass-loss recipe combining \cite{vink01} and \cite{nl00}, as in \cite{mp17}. The evolutionary points at which atmosphere models and synthetic spectra have been computed are indicated with squares.} 
\label{fig:HRDHe}
\end{figure*}

Our new models, including the mass-loss recipe described in Sect.~\ref{sec:massloss}, are in very good qualitative agreement with those of \cite{graef21}: the red bending of the upper the zero-age main sequence (ZAMS), the luminosity decrease along the main-sequence evolution as a result of the huge mass loss they experience, and the envelope inflation is recovered (see also Sect.~\ref{s_spec}). The ZAMS of our models is more blueshifted than Graefener's due to different assumptions for the initial chemical composition. The models with M $\geq$ 200~\msun\ experience optically-thick winds already on the ZAMS, which explains the direct decrease in their luminosity beyond this evolutionary point. The 150~\msun\ model experiences a longer phase with optically thin winds, and thus the slope of its path in the HRD is much more similar to what is obtained when the classical mass-loss recipe is used.

Because all models above 200~\msun\ experience only optically thick winds\footnote{For the 150~\msun\ model, the winds are optically thin down to 53000K and optically thick after that.} , the mass-loss rates predicted by the new mass-loss scheme are high from the ZAMS on, typically a factor 3 to 10 higher than for the optically thin recipe of \citet{vink01} (see e.g. Fig.~8 of \citealt{besten14}).  This leads to an efficient stripping of the stellar envelope, as can be seen from Table~\ref{tab_atm}. This leads to the early exposure of the products of hydrogen nucleosynthesis through the CNO cycle at the surface. Helium is strongly enhanced, as shown in Table~\ref{tab_atm} and Fig.~\ref{fig:HRDHe}. In this figure, we show the evolutionary tracks in the HRD for the models including the \cite{graef21} (left panel) and classical mass loss (right panel). The surface $^4{\rm He}$ mass fraction is colour-coded along the tracks. Comparison between the left and right panels of this figure and with the values reported for the 200 $M_\odot$ models in Tables~\ref{tab_atm} and \ref{tab_atmV} indicates that the helium enrichment is much higher than predicted by the models using the classical mass-loss recipe, with important consequences for the spectral appearance of the low-metallicity VMSs, as we discuss in the following sections.

\section{Spectroscopic properties}
\label{s_spectro}

In this section we first describe the computation of atmosphere models and synthetic spectra. We then discuss the general morphology of our theoretical spectra and compare them to spectra of real stars in the LMC. We also discuss the effect of the new mass-loss rate prescription on synthetic spectra by comparing our results to models using the recipe of \citet{vink01}. 

\subsection{Atmosphere models and synthetic spectra}
\label{s_spec}

\begin{figure*}[ht]
\centering
\includegraphics[width=0.49\textwidth]{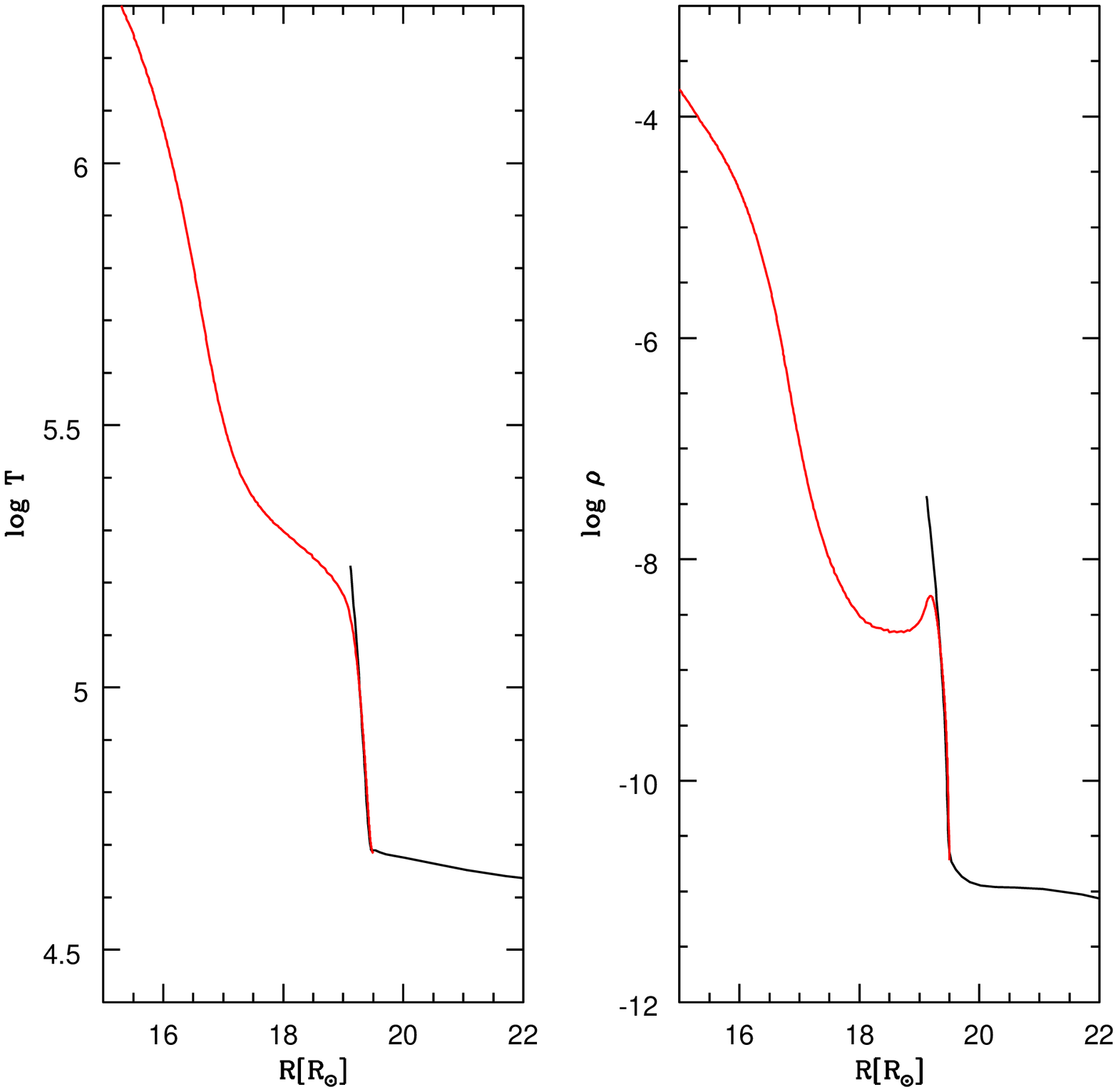}
\includegraphics[width=0.49\textwidth]{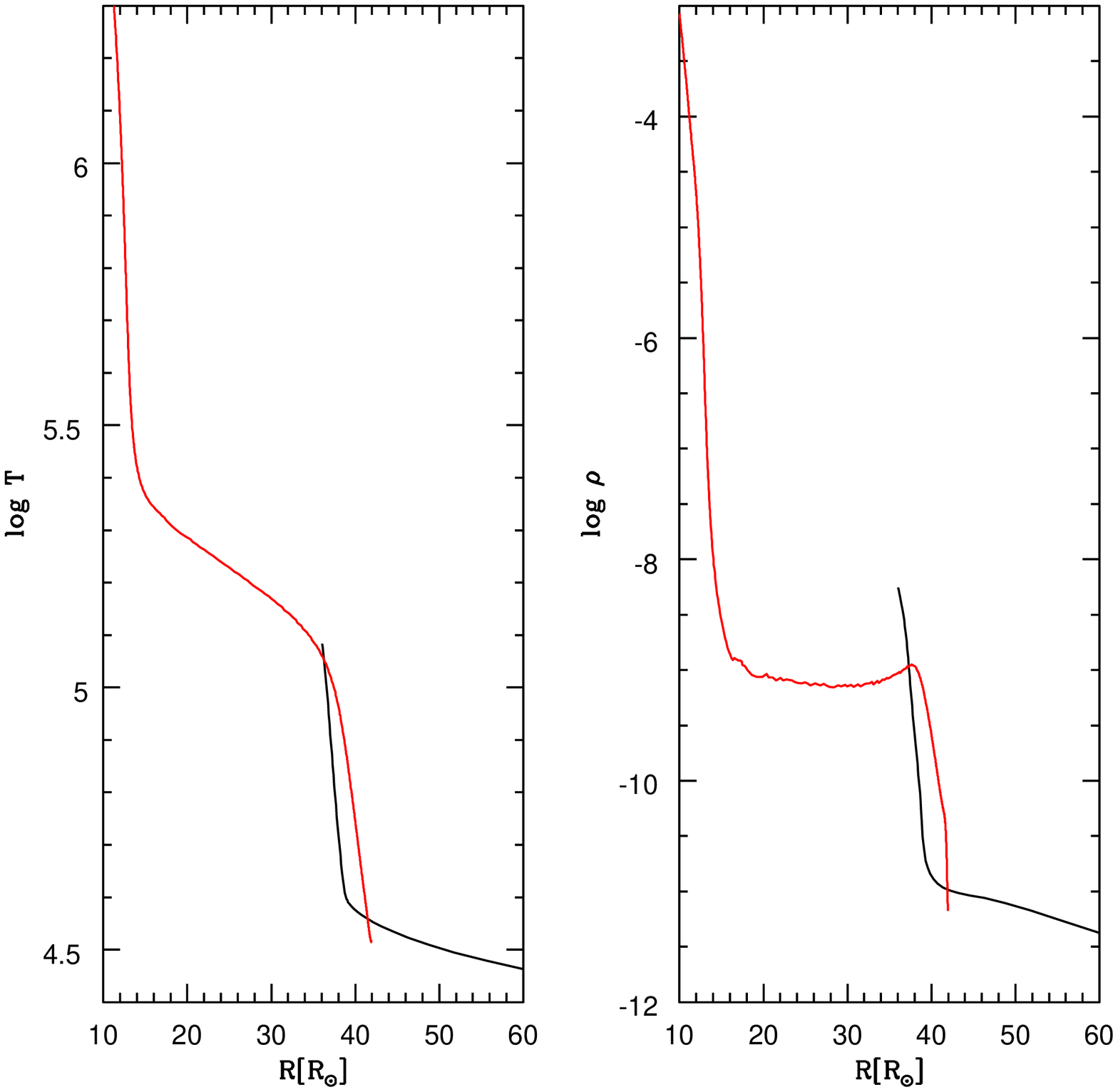}
\caption{Temperature and density structure as a function of radius in models 200a (left) and 200f (right). In each panel, the red line is the structure of the evolutionary model, and the black line is the structure of the atmosphere model.} 
\label{struc200}
\end{figure*}

Along each evolutionary sequence, we selected points for which we computed an atmosphere model and the associated synthetic spectrum with the code \emph{CMFGEN} \citep{hm98}. With applications to population synthesis in mind, we chose points at 0, 0.5, 1.0, 1.5, and 2.0 Myr along each track (when possible, the most massive stars living less than 2.0 Myr). A final point was taken at the redmost extension of the track. Fig.~\ref{fig:HRDHe} shows the positions of the selected points. 

We took the surface properties of the stars predicted by \emph{STAREVOL}  as input for the atmosphere and synthetic spectra calculations. These parameters are gathered in Table~\ref{tab_atm}. The terminal velocity \vinf\ was set to 2.51$\times$\vesc\ as assumed in Sect.\ref{s_mdot}. Because the mass-loss recipe of \citet{graef21} takes wind clumping into account, we also added inhomogeneities in the computation of the atmosphere models. In \emph{CMFGEN,} this is done by means of a volume filling factor $f$ that is assumed to vary as

\begin{equation}
    f = f_{\infty} + (1-f_{\infty})\ e^{-v/v_{cl}}
,\end{equation}

\noindent where $v$ is the wind velocity, $v_{cl}$ is a characteristic velocity at which clumping becomes non-negligible, and $f_{\infty}$ is the maximum clumping factor. We adopted standard values for the last two parameters: $v_{cl}=100$~\kms\ and $f_{\infty}=0.1$ corresponding to $D=10$ in  Eq.~\ref{eq_mdot}.

The following elements were included in the calculations: H, He, C, N, O, Ne, Mg, Si, P, S, Ar, Ca, Fe, and Ni. The abundances were taken from the evolutionary model output. To compute the atmosphere structure, a microturbulent velocity of 20~\kms\ was adopted. For the synthetic spectrum, the microturbulent velocity varied from 10~\kms\ at the photosphere to $0.1\times v_{\infty}$ at the top of the atmosphere. The spectra were calculated from 50~\AA\ to 3~$\mu$m.

Different assumptions are made to compute the temperature and atmosphere structures in the outer regions of evolutionary models and in the inner regions of atmosphere models. It is not obvious that these structures are consistent \citep[see][]{groh14,mp17}. We confirmed this consistency for all our models and found that the agreement is excellent on the ZAMS and degrades in more advanced phases. However, the agreement in our most advanced models is still reasonable. This is illustrated in Fig.~\ref{struc200}, where we took the ZAMS and most evolved models of the 200~\msun\ series (models 200a and 200f) as examples. For model 200a, the internal and atmosphere structures merge almost perfectly. For the 200f model, there is a small offset, but the structures still intersect. We tested that this is no longer the case in more advanced phases. We thus chose not to compute spectra beyond the last model of each track in Fig.~\ref{fig:HRDHe}. This would require a proper treatment of the atmospheric boundary in evolutionary models and is beyond the scope of this study. Similarly, we ran tests for a 500~\msun\ model and found that a strong discrepancy between evolutionary and atmospheric structures is present already on the ZAMS. We thus limited our calculations to masses  $\leq$400~\msun. 

Interestingly, we also note in Fig.~\ref{struc200} that the density structures of our models displays an plateau in the outer regions. This phenomenon is usually referred as "envelope inflation" \citep{ishii99,graef12} and is naturally taken into account in our models. This can be explained by our treatment of subsurface convection in the evolutionary models \citep[see also][]{graef21}.

\begin{table*}[ht]
\begin{center}
\caption{Parameters of the atmosphere models.} \label{tab_atm}
\begin{tabular}{lcccccccccccc}
\hline
model id  &  $T_{\rm eff}$  &  $\log g$ & $\log(L/L_{\odot})$ & R& $\log \dot{M}$ & $\varv_{\infty}$ & age & H & He & C & N & O \\    
  &  K          &                   &          &  R$\odot$ &            & \kms       & Myr  &  & & & & \\
\hline
150 \msun & & & & & & & & & \\
\hline
150a & 57276 & 4.22 & 6.379 &  15.80 & -5.30 &   3641  & 0.01  & 0.738 & 0.257 & 9.741~$10^{-4}$ & 2.852~$10^{-4}$ & 2.362~$10^{-3}$ \\
150b & 53786 & 4.08 & 6.394 &  18.23 & -4.97 &   3239  & 0.50  & 0.738 & 0.257 & 9.741~$10^{-4}$ & 2.852~$10^{-4}$ & 2.362~$10^{-3}$ \\
150c & 49764 & 3.92 & 6.404 &  21.54 & -4.72 &   2856  & 1.01  & 0.738 & 0.257 & 9.741~$10^{-4}$ & 2.852~$10^{-4}$ & 2.362~$10^{-3}$ \\
150d & 44445 & 3.67 & 6.413 &  27.28 & -4.51 &   2280  & 1.50  & 0.709 & 0.286 & 4.841~$10^{-5}$ & 3.284~$10^{-3}$ & 1.685~$10^{-4}$ \\
150e & 40839 & 3.44 & 6.418 &  32.50 & -4.36 &   1858  & 2.00  & 0.539 & 0.456 & 5.111~$10^{-5}$ & 3.408~$10^{-3}$ & 2.256~$10^{-5}$ \\
150f & 38214 & 3.23 & 6.415 &  36.99 & -4.22 &   1503  & 2.50  & 0.251 & 0.744 & 5.538~$10^{-5}$ & 3.405~$10^{-3}$ & 2.050~$10^{-5}$ \\
\hline                                                
200 \msun & & & & & & & & & \\                       
\hline                                                                              
200a & 57476 & 4.16 & 6.564 &  19.41 & -4.71 &   3534  & 0.00  & 0.738 & 0.257 & 9.741~$10^{-4}$ & 2.852~$10^{-4}$ & 2.362~$10^{-3}$ \\
200b & 52830 & 3.99 & 6.564 &  22.98 & -4.59 &   3075  & 0.50  & 0.738 & 0.257 & 6.979~$10^{-4}$ & 6.225~$10^{-4}$ & 2.357~$10^{-3}$ \\
200c & 48281 & 3.80 & 6.566 &  27.57 & -4.42 &   2613  & 1.00  & 0.708 & 0.287 & 4.987~$10^{-5}$ & 3.293~$10^{-3}$ & 1.556~$10^{-4}$ \\
200d & 44687 & 3.60 & 6.565 &  32.15 & -4.30 &   2136  & 1.50  & 0.578 & 0.417 & 5.246~$10^{-5}$ & 3.407~$10^{-3}$ & 2.211~$10^{-5}$ \\
200e & 41127 & 3.37 & 6.558 &  37.65 & -4.19 &   1702  & 2.00  & 0.366 & 0.629 & 5.488~$10^{-5}$ & 3.405~$10^{-3}$ & 2.087~$10^{-5}$ \\
200f & 38781 & 3.18 & 6.539 &  41.38 & -4.11 &   1410  & 2.39  & 0.137 & 0.858 & 5.988~$10^{-5}$ & 3.401~$10^{-3}$ & 1.887~$10^{-5}$ \\
\hline                                                
250 \msun & & & & & & & & & \\                       
\hline                                                                              
250a & 56768 & 4.10 & 6.701 &  23.30 & -4.51 &   3411  & 0.01  & 0.738 & 0.257 & 9.741~$10^{-4}$ & 2.852~$10^{-4}$ & 2.362~$10^{-3}$ \\
250b & 51195 & 3.90 & 6.694 &  28.42 & -4.38 &   2908  & 0.50  & 0.738 & 0.257 & 9.741~$10^{-4}$ & 2.853~$10^{-4}$ & 2.362~$10^{-3}$  \\
250c & 47093 & 3.71 & 6.690 &  33.43 & -4.25 &   2546  & 1.00  & 0.658 & 0.335 & 5.275~$10^{-5}$ & 3.405~$10^{-3}$ & 2.435~$10^{-3}$ \\
250d & 43504 & 3.51 & 6.682 &  38.82 & -4.16 &   2018  & 1.50  & 0.499 & 0.496 & 5.424~$10^{-5}$ & 3.406~$10^{-3}$ & 2.121~$10^{-5}$ \\
250e & 39345 & 3.24 & 6.665 &  46.54 & -4.06 &   1543  & 2.00  & 0.269 & 0.726 & 5.744~$10^{-5}$ & 3.403~$10^{-3}$ & 1.974~$10^{-5}$ \\
250f & 37118 & 3.10 & 6.650 &  51.39 & -4.03 &   1383  & 2.20  & 0.160 & 0.835 & 5.999~$10^{-5}$ & 3.401~$10^{-3}$ & 1.871~$10^{-5}$ \\
\hline                                                
300 \msun & & & & & & & & & \\                       
\hline                                                                              
300a & 55765 & 4.04 & 6.808 &  27.31 & -4.36 &   3306  & 0.01  & 0.738 & 0.257 & 9.741~$10^{-4}$ & 2.853~$10^{-4}$ & 2.362~$10^{-3}$ \\
300b & 49654 & 3.81 & 6.798 &  34.05 & -4.24 &   2711  & 0.50  & 0.729 & 0.266 & 4.397~$10^{-5}$ & 2.775~$10^{-3}$ & 7.551~$10^{-4}$ \\
300c & 45760 & 3.62 & 6.790 &  39.73 & -4.14 &   2265  & 1.00  & 0.620 & 0.375 & 5.385~$10^{-5}$ & 3.406~$10^{-3}$ & 2.149~$10^{-5}$ \\
300d & 42009 & 3.41 & 6.777 &  46.44 & -4.06 &   1935  & 1.50  & 0.442 & 0.553 & 5.562~$10^{-5}$ & 3.405~$10^{-3}$ & 2.055~$10^{-5}$ \\
300e & 37177 & 3.11 & 6.751 &  57.55 & -3.97 &   1444  & 2.00  & 0.209 & 0.786 & 5.938~$10^{-5}$ & 3.402~$10^{-3}$ & 1.893~$10^{-5}$ \\
300f & 35485 & 3.00 & 6.730 &  61.65 & -3.98 &   1313  & 2.14  & 0.134 & 0.861 & 6.153~$10^{-5}$ & 3.400~$10^{-3}$ & 1.813~$10^{-5}$  \\
\hline                                                
400 \msun & & & & & & & & & \\                       
\hline                                                                          
400a & 53023 & 3.92 & 6.972 &  36.49 & -4.16 &   3148  & 0.01  & 0.738 & 0.257 & 9.741~$10^{-4}$ & 2.852~$10^{-4}$ & 2.362~$10^{-3}$ \\
400b & 46762 & 3.67 & 6.957 &  46.11 & -4.05 &   2534  & 0.50  & 0.703 & 0.291 & 5.337~$10^{-5}$ & 3.338~$10^{-3}$ & 9.997~$10^{-5}$ \\
400c & 42744 & 3.46 & 6.944 &  54.36 & -3.97 &   2076  & 1.00  & 0.566 & 0.429 & 5.546~$10^{-5}$ & 3.405~$10^{-3}$ & 2.061~$10^{-5}$ \\
400d & 38183 & 3.19 & 6.926 &  66.73 & -3.90 &   1613  & 1.50  & 0.372 & 0.623 & 5.760~$10^{-5}$ & 3.403~$10^{-3}$ & 1.964~$10^{-5}$ \\
400e & 35065 & 3.00 & 6.910 &  77.68 & -3.86 &   1393  & 1.77  & 0.250 & 0.745 & 5.958~$10^{-5}$ & 3.402~$10^{-3}$ & 1.884~$10^{-5}$ \\
\hline
\end{tabular}
\tablefoot{Columns are model number, effective temperature, surface gravity, luminosity, radius, mass-loss rate, wind terminal velocity, age, and surface abundances of H, He, C, N, and O given in mass fraction. For each mass, the models are ordered by increasing age (model with an a index are the ZAMS model, and models with an e or f index are the last point along the evolutionary track in the HRD; see Fig.~\ref{fig:HRDHe}).}
\end{center}
\end{table*}

\subsection{Spectral morphology}
\label{s_morpho}

We illustrate the evolution of the spectroscopic appearance of our VMS models with the 200~\msun\ star in Fig.~\ref{sv200}. The figures for the other masses are gathered in Appendix~\ref{ap_spec}. 
To better identify spectral lines, we show in Fig.~\ref{lineid} the contribution of individual ions to the total UV spectrum of model 200d and to the optical spectrum of model 200f. These two models were chosen because they allow the identification of the maximum number of individual lines in the UV and optical range, respectively. We distribute our synthetic spectra through the \emph{POLLUX} database\footnote{\url{pollux.oreme.org/}} \citep{pollux}. 

In the next sections we describe the morphology of our VMS spectra in the optical and UV range. We highlight the features that appear as typical of VMSs and are not observed in stars with M~$\lesssim$~100~\msun\ .

\begin{figure*}[ht]
\centering
\includegraphics[width=0.49\textwidth]{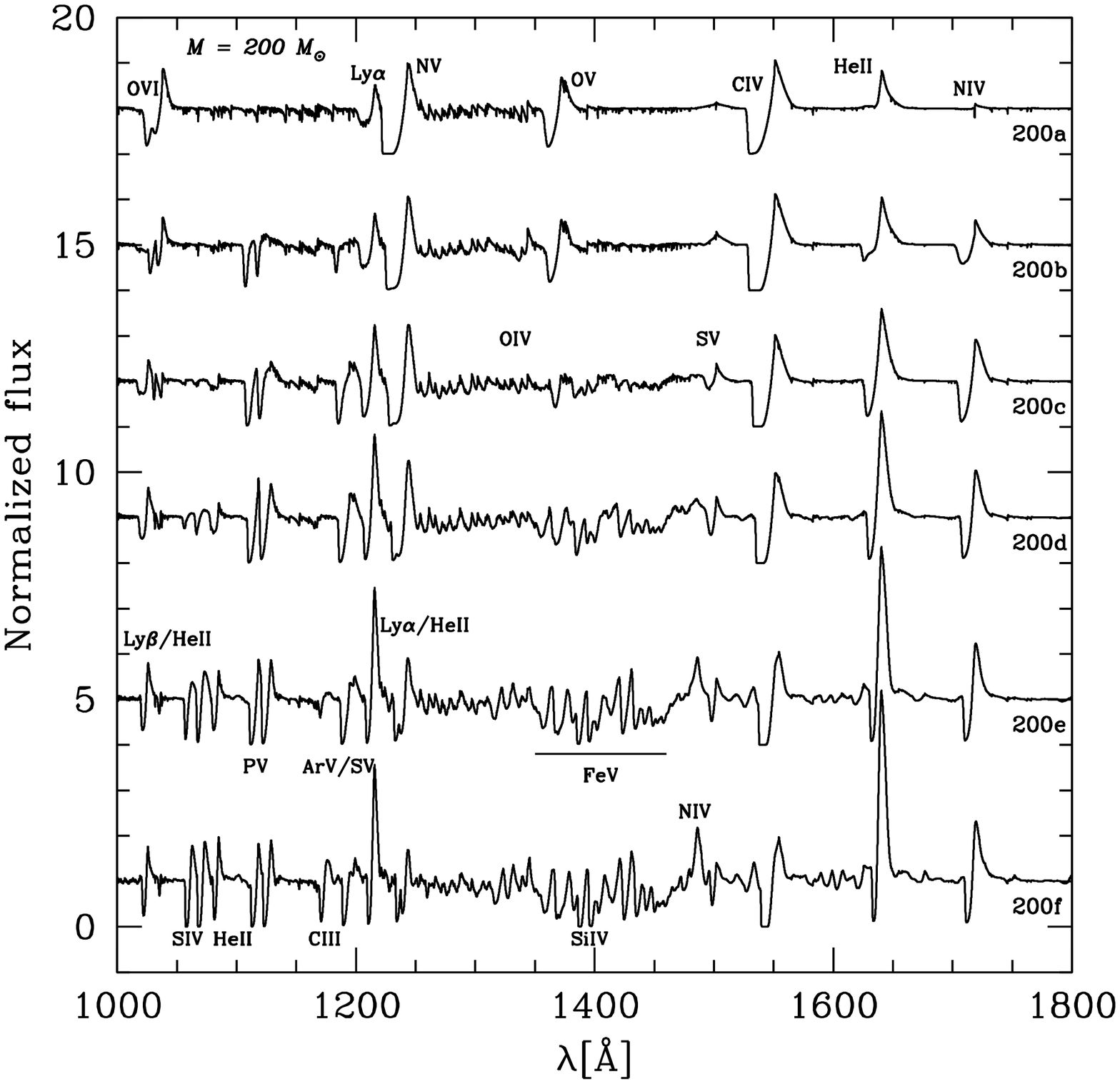}
\includegraphics[width=0.49\textwidth]{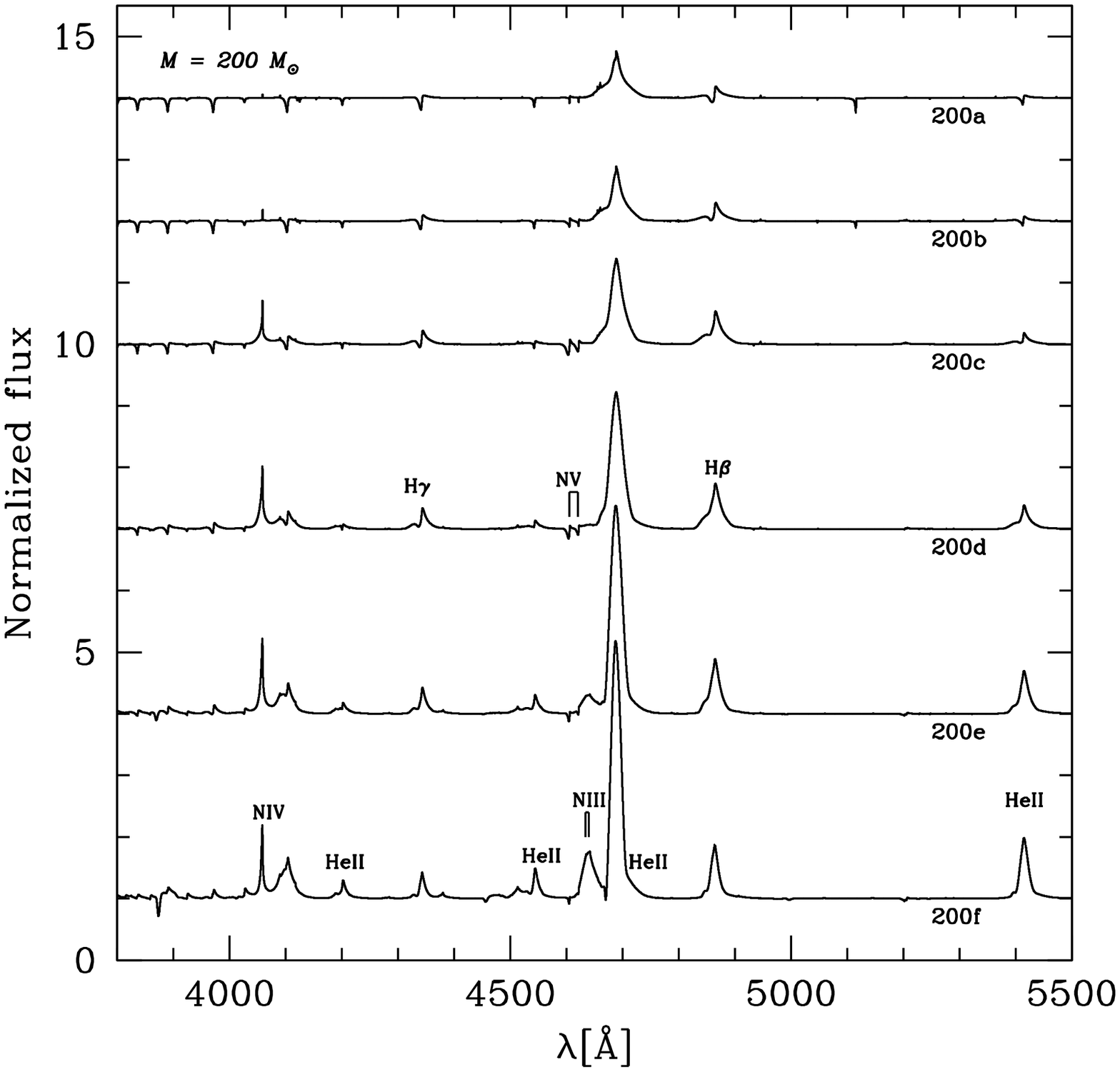}
\caption{UV (left panel) and optical (right panel) synthetic spectra of the M=200~\msun\ series. The spectra are normalized and shifted upward for clarity. The main lines are indicated. Evolution proceeds from top to bottom.} 
\label{sv200}
\end{figure*}

\subsubsection{Optical range}
\label{s_opt}

\begin{figure*}[ht]
\centering
\includegraphics[width=0.49\textwidth]{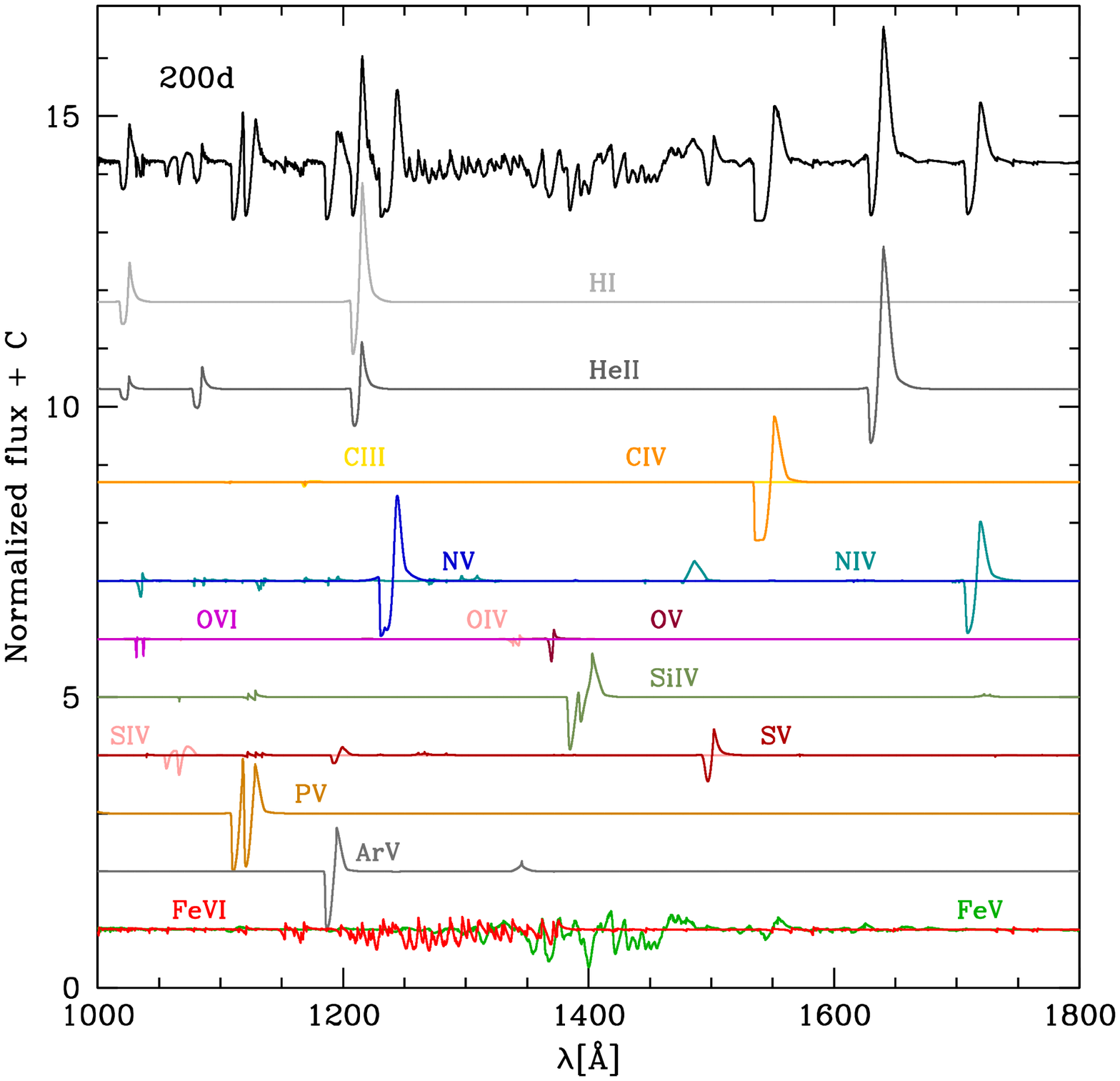}
\includegraphics[width=0.49\textwidth]{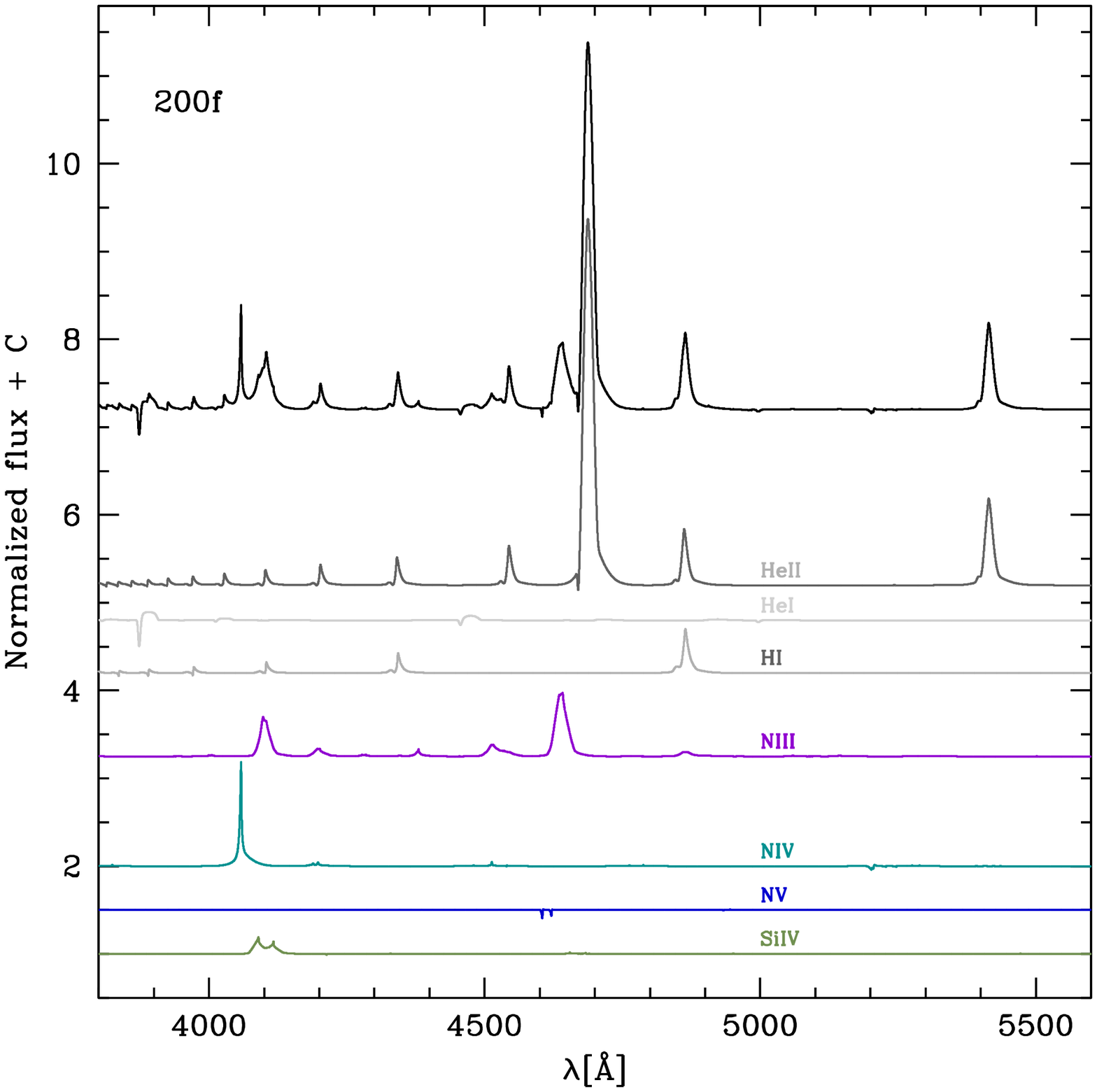}
\caption{UV spectrum of the 200d model (left) and optical spectrum of the 200f model (right). In each panel, the black line is the total spectrum, and coloured lines indicate the contribution of individual ions.}
\label{lineid}
\end{figure*}

In the optical range, it is immediately clear that the spectra are dominated by \heiiopt\ emission that increases as the star evolves. For the 200~\msun\ models, the EW of this emission ranges from -9 to -55~\AA\footnote{We adopt the usual convention of negative EW for emission.}. Other \ion{He}{ii} lines are predicted in absorption near the ZAMS and in emission in more advanced phases. Hydrogen lines of the Balmer series are seen either in absorption or in full emission, depending on the line and the state of evolution. In more evolved models, the contribution of \ion{H}{i} lines weakens because the hydrogen content is reduced (see Table~\ref{tab_atm}). This implies that the emission features at the position of the Balmer hydrogen lines become dominated by \ion{He}{ii} lines as evolution proceeds; see for instance the emission at the position of H$\beta$ in model 200f (Fig.~\ref{lineid}, right panel).

Nitrogen lines from \ion{N}{iii}, \ion{N}{iv,} and/or \ion{N}{v} are the next strongest features in the optical range. Their relative strength varies depending on the ionization of the atmosphere: \ion{N}{v} is present at the highest temperatures and recombines into \ion{N}{iv} and \ion{N}{iii} when \teff\ decreases and the wind density increases. For instance, \ion{N}{v}~4605-20 and \ion{N}{iv}~4058 are predicted in model 200d, where \ion{N}{iii}~4640 is very weak. In model 200f, \ion{N}{v}~4605-20 has almost vanished, while \ion{N}{iii}~4640 is comparable to \ion{N}{iv}~4058. In addition to ionization effects, the rapid increase in nitrogen content at the surface of the stars causes a global strengthening of all nitrogen lines. Wind effects also play a role \citep{rivero11,rivero12}: the presence of velocity fields leads to a desaturation of some key transitions that control the populations of the energy levels of optical lines. Consequently, the strength of these lines strongly depends on the wind properties. We stress that \ion{N}{iii} is the main contributor to the emission complex located near the position of H$\delta$ in model 200f (and similarly evolved models). \ion{H}{i}~4101 and \ion{Si}{iv}~4089-4116 emissions are present, but much weaker than \ion{N}{iii}~4097. 

These spectral properties are typical of either O supergiants with strong winds (OIf stars), transition objects (OIf/WN), or hydrogen-rich WN (WNh) stars, as witnessed by the emission in \heiiopt\ \citep{martins18,ssm96}. The morphology of H$\beta$ can be used to further distinguish between early OIf, OIf/WN, and WNh stars, as demonstrated by \citet{crowal11}: H$\beta$ is in absorption, with a P-Cygni profile, and in emission at these spectral types. Near the ZAMS, none of our models shows \ion{He}{i} lines. Together with the H$\beta$ morphology, this indicates a spectral type O2-3.5If/WN5-7 for the early phases of the 200~\msun\ star evolution. As \teff\ decreases and the wind strengthens, the spectral appearance shifts to WN6-7h. This trend is present for other masses and is modulated by the wind strength. The 150~\msun\ series has a lower wind density on average than the 200~\msun\ series, implying a start as O2-3.5If instead of O2-3.5If/WN5-7. 

We predict a morphology of optical lines that is typical of VMSs observed in 30~Dor, for instance (see \citealt{besten20} and Sect.~\ref{s_realstars}). Less massive O stars on the main sequence do not show these emission features \citep{sota11,sota14}. \heiiopt\ is usually in absorption, and none of the hydrogen Balmer lines shows emission (except sometimes H$\alpha$). Additional examples of LMC O star optical spectra can be found in the supplementary material of \citet{besten20}.

\subsubsection{UV range}
\label{s_uv}

The typical UV spectrum of our VMS models is made of several P-Cygni profiles originating from different ions (Fig.~\ref{lineid}, left panel). 
As in the optical range, \ion{He}{ii} lines are relatively strong. \heiiuv\ starts as an emission line on the ZAMS (model 200a) and develops rapidly (from model 200c) into the strongest P-Cygni profile of the UV range as the star evolves (see Fig.~\ref{sv200}). The increase in wind density and of the surface helium content causes this behavior. 
With the exception of the first two models of the 150~\msun\ series (Fig.~\ref{sv150}), \heiiuv\ is always at least comparable in strength to \civuv, and often stronger. Its full width at half maximum is about 1000~\kms. Its EW can reach up to -30~\AA. These properties make \heiiuv\ a unique feature of VMSs on the main sequence. 

\civuv\ is present in all spectra with a prominent P-Cygni profile.  Fig.~\ref{sv200} shows that its strength remains roughly constant along the sequence. The expected increase due to the higher wind density is counter-balanced by the reduction of the carbon surface abundance by more than a factor of 10 between models 200a and 200f. The EW of the absorption (emission) part of \civuv\ is 14 (-10~\AA) and 9 (-9~\AA) in these two models, respectively.

\nivuv\ is weak near the ZAMS, but rapidly develops into a strong P-Cygni profile, mainly because of the fast nitrogen enrichment of the surface (increase by a factor of 10 between models 200a and 200c). The intensity of \nivuv\ is comparable to or even exceeds that of \civuv. 

\ion{N}{iv}~1486 is predicted in emission towards the end of the series. It becomes one of the strongest features of VMS spectra. It is not observed in 40-85~\msun\ O-type supergiants with strong winds (OIf stars) even at solar metallicity \citep{walborn85,bouret12}. However, it is a rather common feature of WN and WNh stars \citep{hamann06,hainich14}.

At the bottom of Fig.~\ref{lineid}, the \ion{Fe}{v} (and to a lesser extent, \ion{Fe}{vi}) line forests shape the morphology of the spectrum around 1400~\AA\ (1300~\AA). These forests appear as broad features in the most advanced phases of our models. The \ion{Fe}{v} forest reduces the emission peak of the
 \ion{Si}{iv}~1400 doublet, which would otherwise appear as a well-developed double P-Cygni profile. 

Finally, \ion{Ar}{v}~1194 is a strong line partly blended with the weaker \ion{S}{v}~1188 feature. Its proximity to \lya\ makes it difficult to actually observe because of broad interstellar absorption (as seen in Fig.~\ref{r136r146} and \ref{r136all}). Nonetheless, to the best of our knowledge, this feature is not mentioned in atlases of O and Wolf-Rayet stars and might be a metallicity indicator in low-extinction regions. 

\smallskip

In view of this general morphology, we conclude that a strong \heiiuv\ emission is a clear signature of VMSs. The presence \ion{N}{iv}~1486 is another specific feature, and \nivuv\ can also be associated with VMSs when its strength is comparable to that of other P-Cygni profiles in the UV range.

\subsection{Comparison to LMC stars}
\label{s_realstars}

\begin{figure}[t]
\centering
\includegraphics[width=0.49\textwidth]{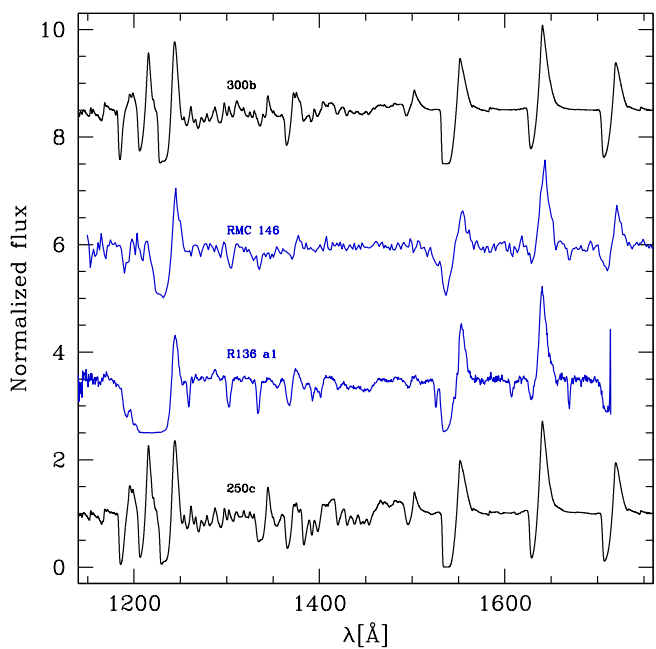}
\caption{UV spectrum of the WN5h stars R136~a1 and RMC~146 (blue middle spectra) with our models 250c and 300b.}
\label{r136r146}
\end{figure}

To determine the relevance of our synthetic spectra for representing the general morphology of VMSs at a metallicity representative of the LMC, we compare them with the normalized spectrum of the WN5h star R136~a1 in Fig.~\ref{r136r146}. The data are from \citet{crowther16} (see their Fig.~B1). According to \citet{besten20}, R136~a1 has a luminosity of $10^{6.79\pm0.10}$ L$_{\odot}$. Hence we overplot the spectra of our 300~\msun\ and 250~\msun\ series that resemble the spectrum of R136~a1 most. The morphology of the synthetic spectra is similar to that of the observed spectrum. The strong P-Cygni profiles of \ion{N}{v}~1240 and \ion{C}{iv}~1550 are well accounted for. \heiiuv\ is the strongest line in models 300b and 250c, as observed for R136~a1. \ion{S}{v}~1500 and \ion{O}{v}~1371 are predicted as too strong in our models. \ion{Si}{iv}~1400 is also present in our models.

Another sanity check was performed with the WN5h star RMC~146 in the LMC\footnote{The observed spectrum of RMC~146 is SWP04140 in the IUE database of the MAST archive ({\em  https://archive.stsci.edu/iue/search.php}).}. \heiiuv\ is observed as the strongest emission line as in models 300b and 250c. \civuv\ and \nivuv\ have comparable emission peaks, as in the models, although the absolute emission level is slightly lower than predicted. For \nivuv,\ this can partly be due to a different initial nitrogen content: we used a scaled solar value, which is probably higher than the baseline LMC nitrogen abundance \citep{dopita19}. \ion{S}{v}~1500 and \ion{Si}{iv}~1400 are not observed in RMC~146. \ion{Si}{iv}~1400 is sensitive to \teff\ (and thus age), and so is \ion{O}{v}~1371, which is observed to be weaker than predicted in model 300b. 

Since we did not aim at a detailed fit of these two stars but at a morphological comparison, we conclude that our models are satisfactory to account for the general UV features seen in R136~a1 and RMC~146. When optical data for these objects are publicly available, we will perform a similar comparison with our models. A preliminary qualitative inspection of Fig.~S12 of \citet{besten20} indicates that our synthetic spectra look quite similar to the observed spectrum of R136~a1.

\begin{table*}[ht]
\begin{center}
\caption{Same as Table~\ref{tab_atm} for the 200~\msun\ models computed with the Vink et al. mass-loss rate recipe.} \label{tab_atmV}
\begin{tabular}{lcccccccccccc}
\hline
model id  &  $T_{\rm eff}$  &  $\log g$ & $\log(L/L_{\odot})$ & R & $\log \dot{M}$ & $\varv_{\infty}$ & age & H & He & C & N & O \\    
  &  K          &                   &          &  R$\odot$ &             & \kms       & Myr  &  & & & & \\
\hline
200Va & 57183 & 4.15 & 6.563 &  19.59 & -5.07 &   3504  & 0.00  & 0.738 & 0.257 & 9.747~$10^{-4}$ & 2.860~$10^{-4}$ & 2.361~$10^{-3}$ \\
200Vb & 52774 & 3.99 & 6.579 &  23.43 & -4.94 &   3112  & 0.50  & 0.738 & 0.257 & 8.568~$10^{-4}$ & 4.363~$10^{-4}$ & 2.358~$10^{-3}$ \\
200Vc & 47288 & 3.76 & 6.598 &  29.82 & -4.82 &   2553  & 1.01  & 0.738 & 0.257 & 6.571~$10^{-4}$ & 7.100~$10^{-4}$ & 2.311~$10^{-3}$ \\
200Vd & 38032 & 3.34 & 6.621 &  47.34 & -4.79 &   1859  & 1.51  & 0.727 & 0.268 & 1.818~$10^{-4}$ & 2.291~$10^{-3}$ & 1.128~$10^{-3}$ \\
200Ve & 23507 & 2.46 & 6.650 & 128.13 & -4.24 &   1032  & 2.00  & 0.690 & 0.305 & 5.121~$10^{-5}$ & 3.378~$10^{-3}$ & 5.667~$10^{-5}$ \\
\hline    
\end{tabular}
\end{center}
\end{table*}

\subsection{Comparison to Vink et al.'s mass-loss rates}
\label{s_Vink}

\begin{figure*}[t]
\centering
\includegraphics[width=0.49\textwidth]{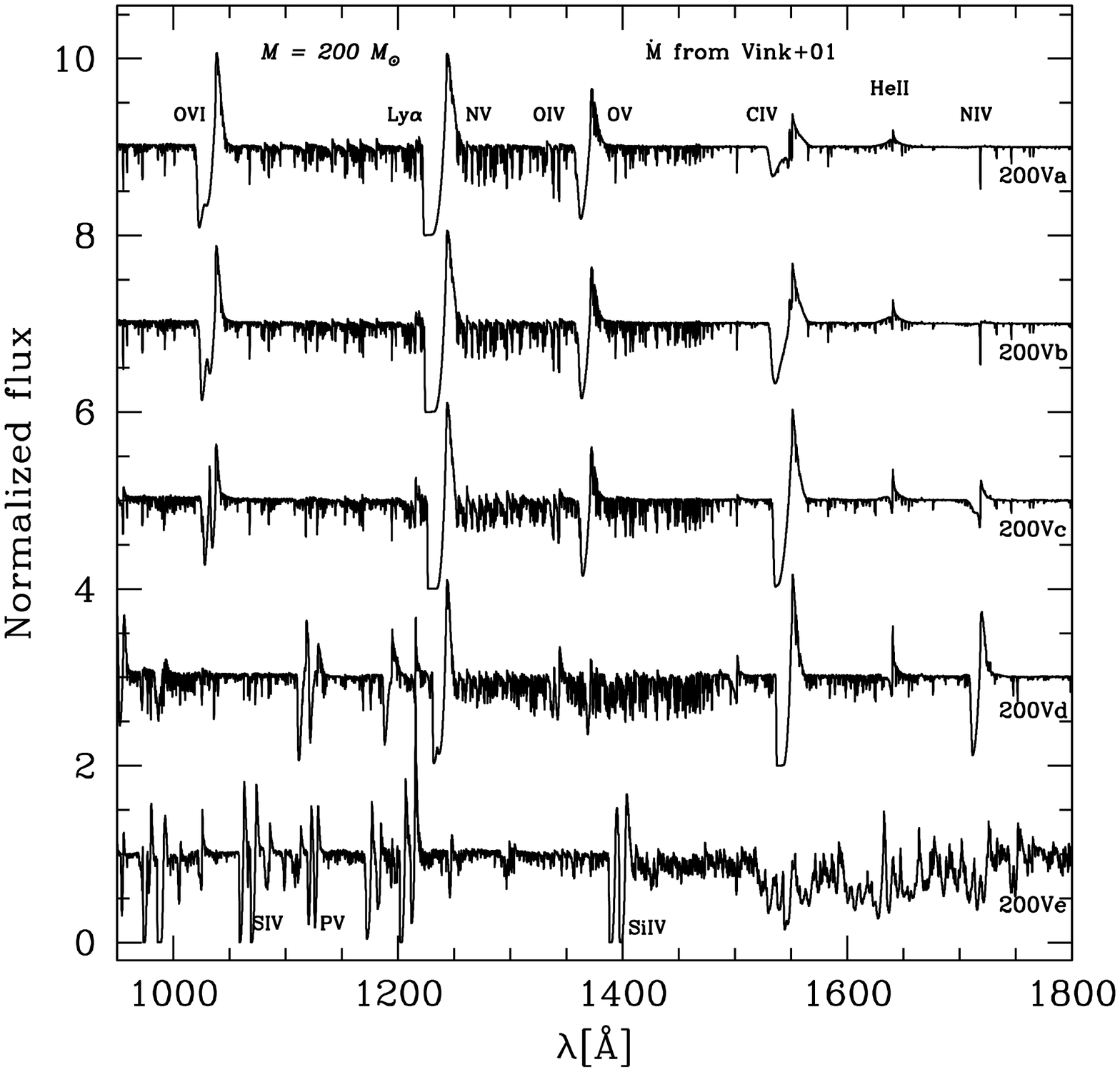}
\includegraphics[width=0.49\textwidth]{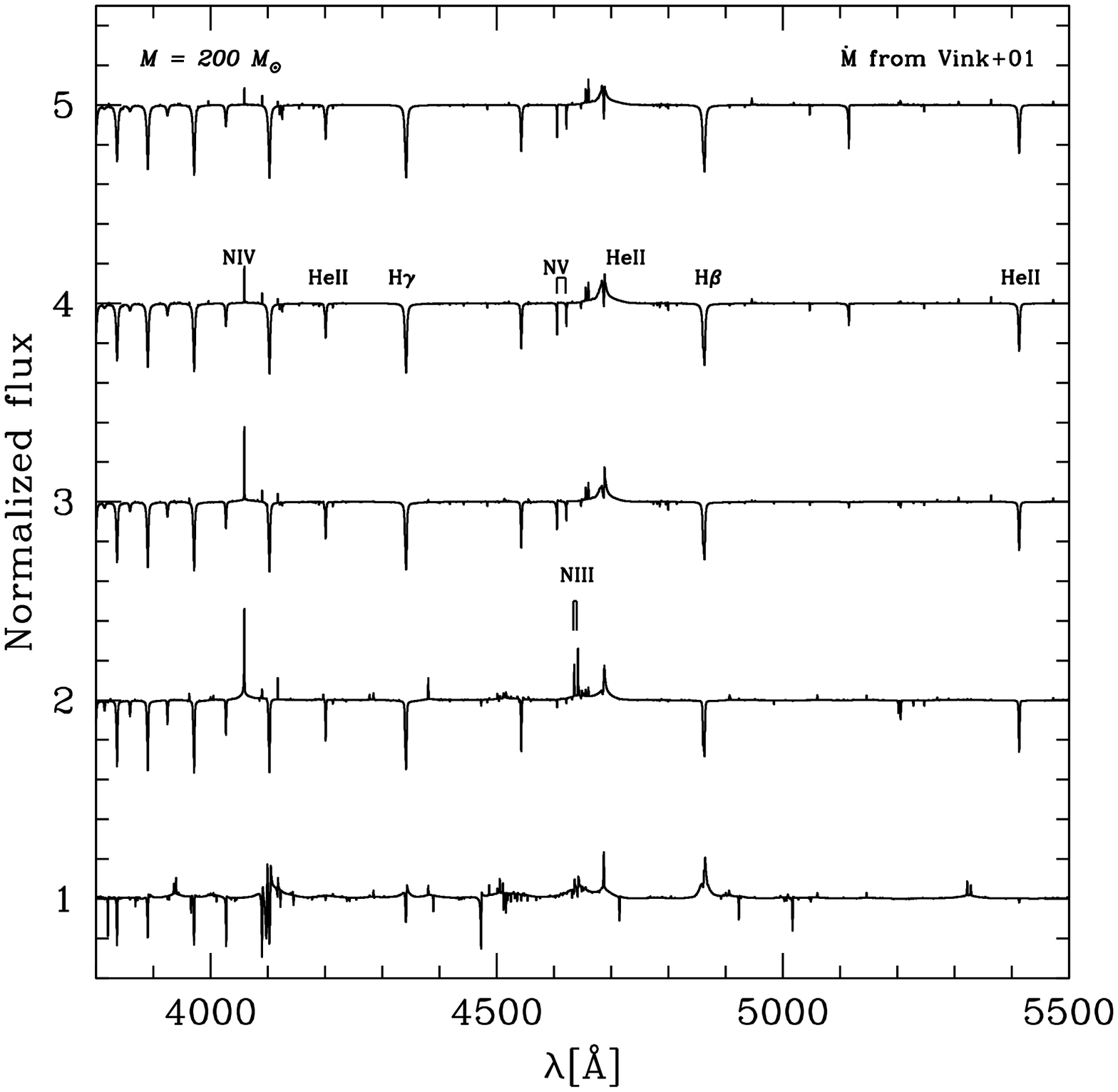}
\caption{Same as Fig.~\ref{sv200}, but for models with the \citet{vink01} mass-loss rate recipe.} 
\label{sv200V}
\end{figure*}

The effect of the new mass-loss rate recipe on the spectra of VMSs can be assessed from Table~\ref{tab_atmV} and Fig.~\ref{sv200V}. We gathered the parameters of models computed with the mass loss prescription of \citet{vink01} for the 200~\msun\ series (200V series in the following). For these calculations, we used the 200~\msun\ track presented in the right panel of Fig.~\ref{fig:HRDHe}. We stress that these computations do not include clumping. In the computations using the mass-loss recipe of \citet{graef21}, clumping is taken into account for optically thick winds, but not for optically thin winds, see Sect.~\ref{s_compev}. Clumping would have affected mostly the red wing of some emission lines (e.g. \heiiopt\ and Balmer lines; see  \citealt{hil91,martins09}). It would also have improved the predicted shape of some UV features, in particular, \ion{P}{v}~1118-28, \ion{O}{v}~1371, and \ion{N}{iv}~1720 \citep{bouret05,fullerton06}. For our computations, we chose the same ages as in the original series to compute the synthetic spectra. Due to the reduced mass-loss rates, the 200~\msun\ track extends towards lower \teff\ (see Fig.~\ref{fig:HRDHe}). 

With the Vink et al. mass-loss rates, \heiiuv\ does not develop into a strong P-Cygni profile and remains relatively weak compared to other UV lines (Fig.~\ref{sv200V}, left panel). Its EW (emission part) barely reaches -1.5~\AA\ in model 200Vd. The same conclusion applies to \heiiopt\ (emission EW smaller than 1.2~\AA). This is due to a combination of a lower mass-loss rate and a lower helium content. Table~\ref{tab_atmV} shows that \mdot\ is between 0.1 and 0.6 dex lower in the 200V series than in the initial 200 models. The helium mass fraction never exceeds 0.3 in the 200V models, while it reaches 0.85 in the initial models, as also illustrated in Fig.~\ref{fig:HRDHe}.

In contrast to the initial 200~\msun\ sequence, \civuv\ increases along the 200V sequence, except for the last model, in which the low \teff\ implies a low \ion{C}{iv} fraction. This trend is explained by the increase in wind density and a relatively constant surface carbon abundance, at least in models 220Va, 200Vb, and 200Vc. Only in model 200Ve, at 2~Myr, is the drop in abundance sufficient to counterbalance the wind effect. 
In the initial series, the C abundance drop takes place much earlier, before 1~Myr. 

With the Vink et al. mass-loss rates, the 200~\msun\ track extends towards lower \teff\ where the \ion{Si}{iv}~1400 doublet can develop into strong P-Cygni profiles. In the initial series, \teff\ is higher, and as described previously, the \ion{Fe}{v} line forest develops into relatively broad features that reduce the \ion{Si}{iv}~1400 lines. 

Another key difference is the shape of the iron line forests. With the weaker mass-loss rates implied by the Vink et al. recipe, these lines are formed closer to the photosphere where wind velocity is low. Hence, lines do not overlap because of Doppler shifts. This is different when the stronger mass-loss rates of Graefener are used. In this case, lines are formed higher in the atmosphere where velocities imply Doppler shifts comparable to the intrinsic line width. Hence blending creates the broad absorption features seen in the left panel of Fig.~\ref{sv200}, for example. 

The spectrum of VMSs is thus very different depending on the mass-loss rate prescription. Comparisons to LMC stars (see previous section) revealed that models need to produce \heiiuv\ at least as strong as \civuv\ to correctly account for the observed UV morphology. Inspection of Fig.~\ref{sv200V} shows that this is not achieved with the Vink et al. recipe. Our study thus favours the recipe of \citet{graef21} to model VMSs.

\section{Effects on population synthesis}
\label{s_popsyn}

\begin{figure*}[t]
\centering
\includegraphics[width=0.49\textwidth]{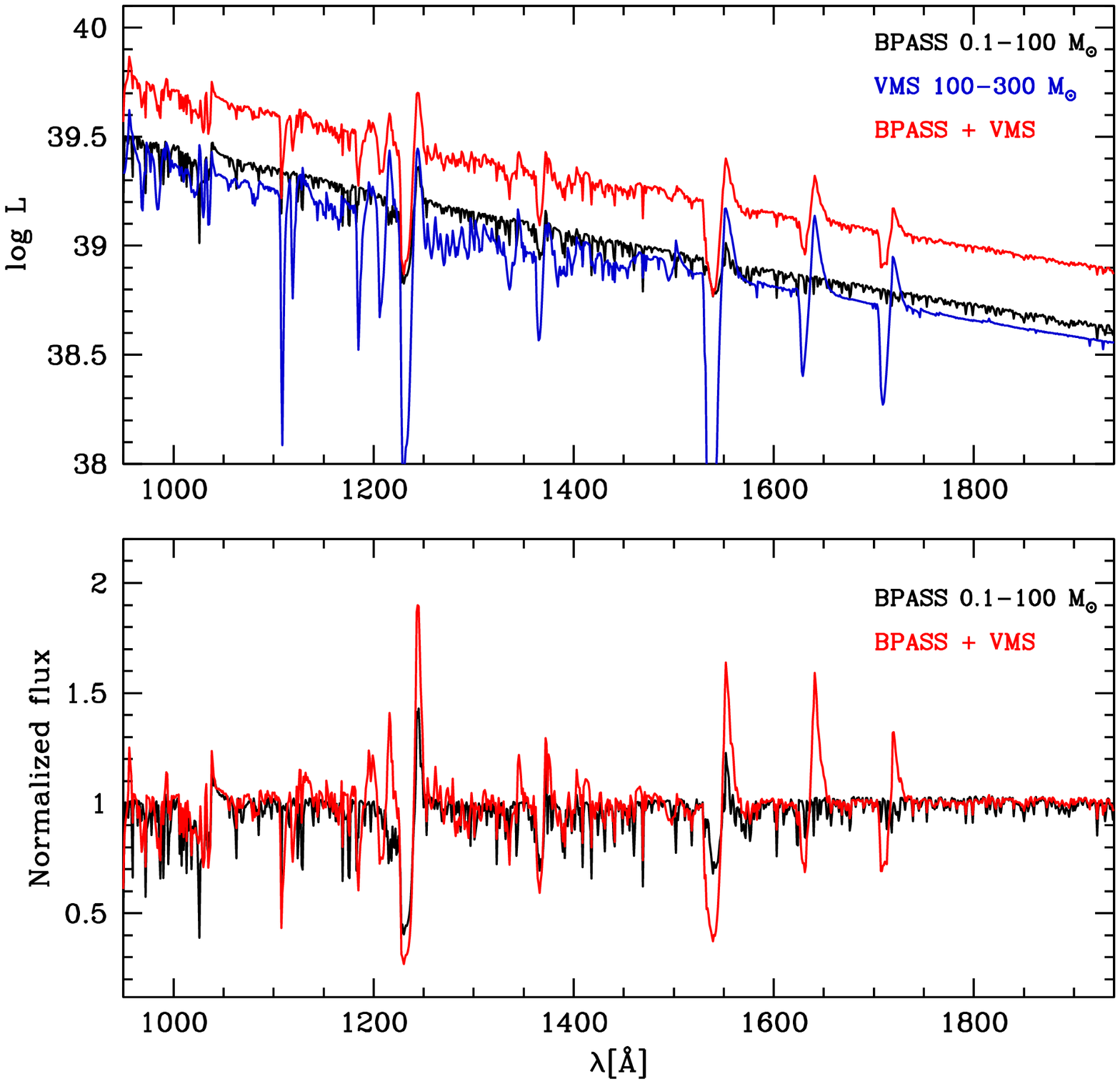}
\includegraphics[width=0.49\textwidth]{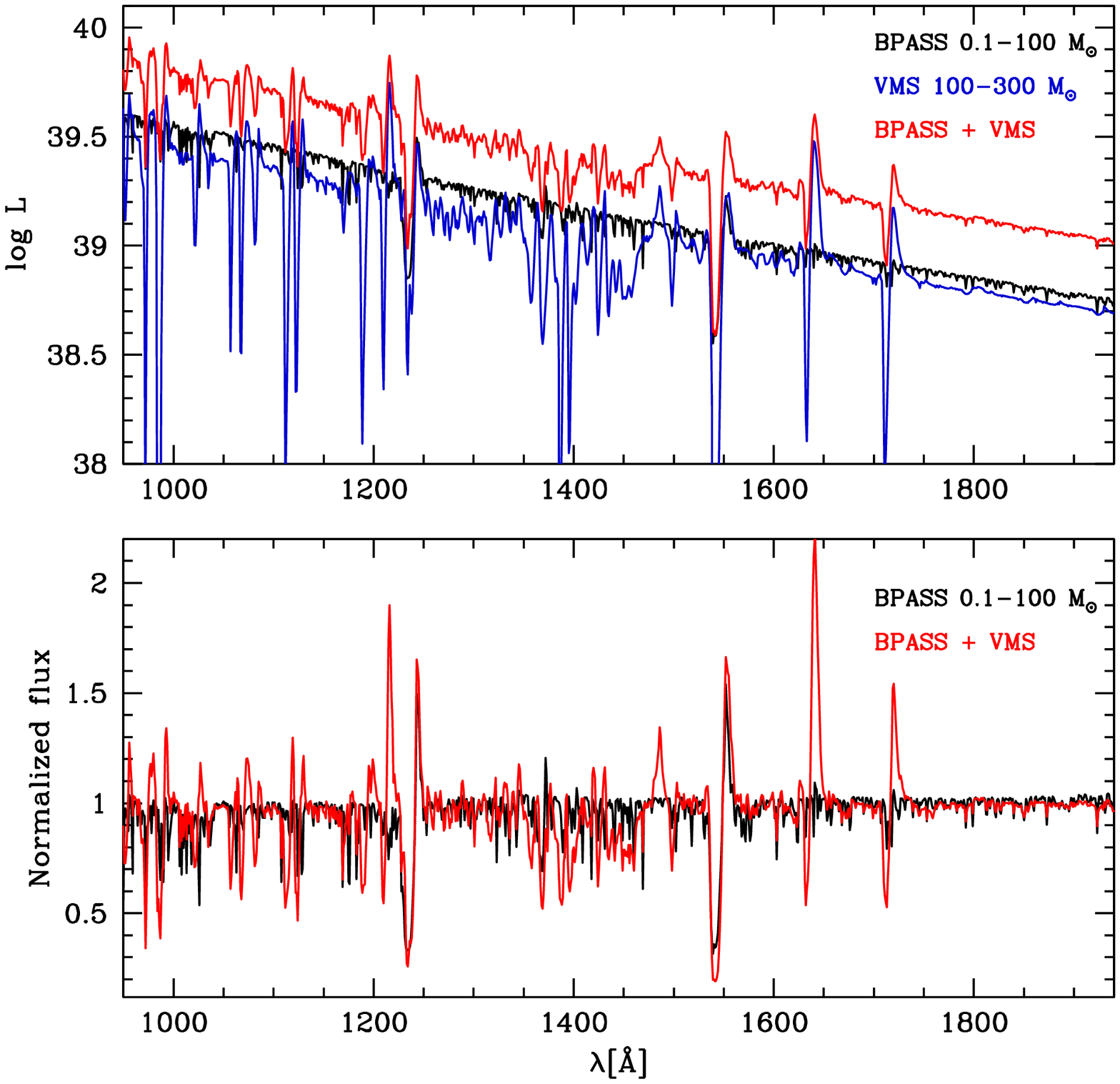}
\caption{UV spectrum of single-star \emph{BPASS} population synthesis models at 1~Myr (left panel) and 2~Myr (right panel) shown in black. The blue spectrum is the contribution of stars with masses in the range 100-300~\msun. The red line is the spectrum of the entire population. In each panel the comparison between the normalized spectrum of the population with (red) and without (black) VMS is shown at the bottom.}
\label{bpass}
\end{figure*}

In this section we investigate the impact of VMS spectra on population synthesis models incorporating stars up to 300~\msun. We show that a proper treatment of stellar evolution and stellar atmospheres with the mass-loss recipe relevant for VMSs has a strong impact on the appearance of young stellar populations. We first discuss theoretical models and how we add VMSs, before making comparisons to young massive clusters.

\subsection{Theoretical predictions}
\label{s_popsyntheo}

To build population synthesis models incorporating VMSs, we retrieved the spectral energy distribution (SED) of the \emph{BPASS} single-star population synthesis models\footnote{\url{https://bpass.auckland.ac.nz/}} for a metallicity Z=0.006, a Salpeter IMF above 0.5~\msun, and an instantaneous burst of star formation. The metallicity is the closest to Z~=~1/2.5~\zsun\ in the \emph{BPASS} models. We refer to \citet{bpass} for a description of the \emph{BPASS} models.

We first used \emph{BPASS} models that extend up to 100~\msun. To assess the effect of VMSs on population synthesis models, we added to these \emph{BPASS} models the contribution of stars in the mass range 100-300~\msun. For this, we used the same mass function as in the \emph{BPASS} calculations and added the contribution of stars in the mass bins 100-175, 175-225, 225-275, and 275-300~\msun\ as detailed in Appendix~\ref{ap_imf}. For each mass bin, we used the SEDs of the 150, 200, 250, and 300~\msun\ series. We focused on two ages -- 1 and 2~Myr -- because VMSs rapidly disappear after this. 

Fig.~\ref{bpass} shows the results of this exercise in the UV wavelength range. VMSs contribute almost the same amount of light as all 0.1-100~\msun\ stars. Some lines are almost exclusively produced by VMSs. This includes \heiiuv\ and \nivuv. \ion{P}{v}~1117-28, \ion{C}{iv}~1169, \ion{N}{v}~1240, and \ion{C}{iv}~1550 are also affected, they are stronger when VMSs are included, especially at 2~Myr. At this age, \ion{N}{iv}~1486 appears in the integrated spectrum because of VMSs.

\begin{figure}[t]
\centering
\includegraphics[width=0.49\textwidth]{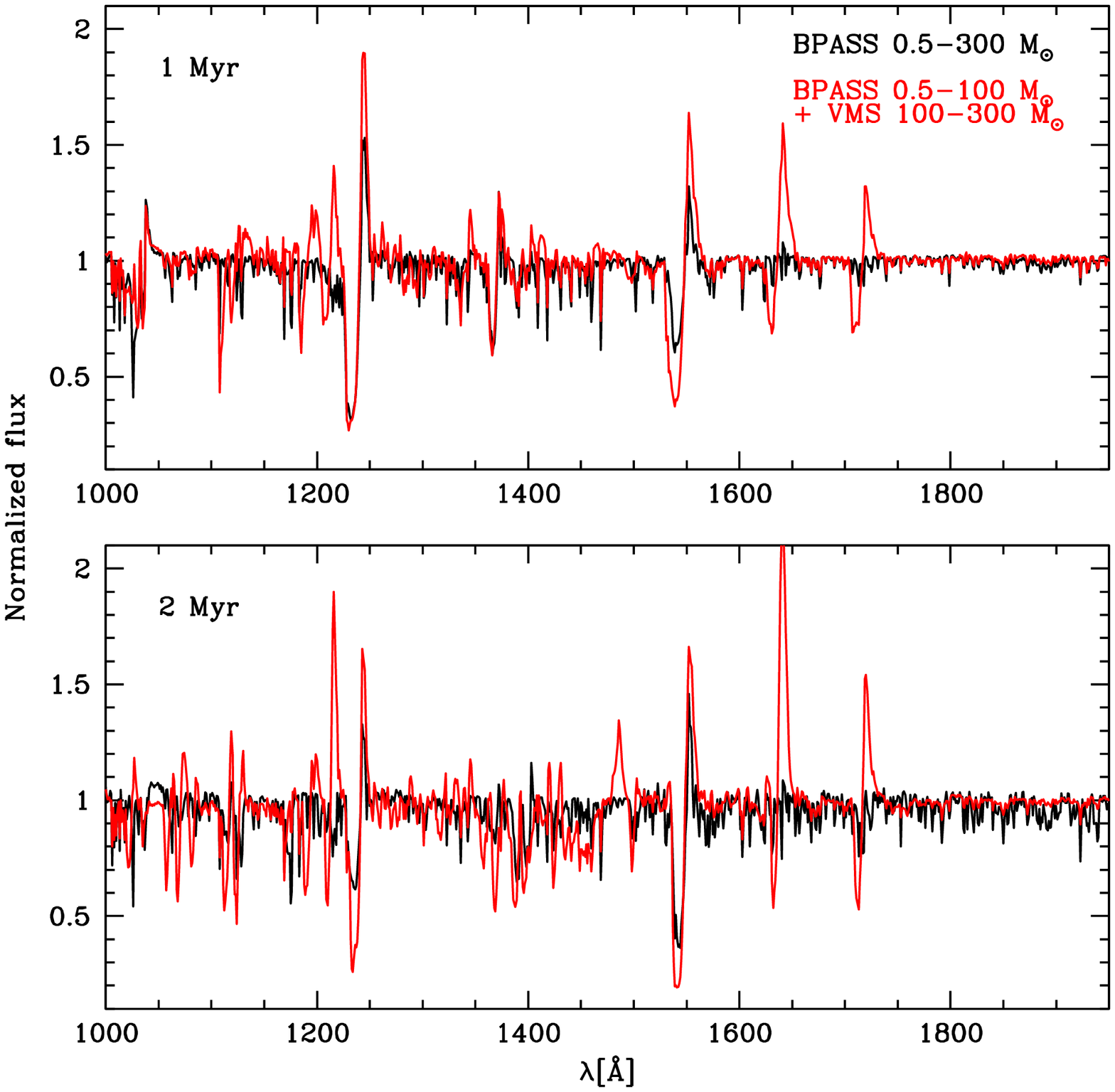}
\caption{Normalized UV spectrum of population synthesis models. Black lines are the  \emph{BPASS} model extending to 300~\msun. The red line is the combination of the \emph{BPASS} model up to 100~\msun\ and the additional contribution of VMSs from our models and a Salpeter IMF slope for 100-300~\msun. The top (bottom) panel is for a burst of star formation after 1~(2)~Myr.}
\label{bpass300}
\end{figure}

The \emph{BPASS} database provides models that include stars up to 300~\msun. For this purpose, both the evolutionary tracks and the stellar atmospheres for 100-300~\msun\ stars were computed with the mass-loss rates of \citet{vink01}.  This causes differences compared to population synthesis models based on the appropriate treatment of VMS winds. We illustrate this in Fig.~\ref{bpass300}. All wind lines are stronger when proper VMS mass-loss rates are included. This is again striking for \heiiuv\ and \nivuv. The former is produced exclusively when our VMS models are taken into account. \civuv\ is also affected, being stronger in our predictions. At 2~Myr \ion{N}{iv}~1486 is present only in our models. The strength of all lines affected by VMSs depend on the number of such objects. We recall that in our tests we extended the Salpeter mass function up to 300~\msun. A different IMF slope and/or a different upper mass cut-off would modulate the shape of all strong wind-sensitive lines. Our tests are also made for a burst of star formation, and further studies should investigate various star formation histories.

\citet{senchyna21} produced improved population synthesis models with the Bruzual \& Charlot code \citep{vidal17}. Unlike the \emph{BPASS} models, they include new evolutionary tracks from \citet{chen15} that take the specific mass-loss rates of VMSs into accoumt. The mass-loss recipe used by Chen et al. is that of \citet{vink11} and is slightly different from that of \citet{graef21}. In particular, the scaling of the mass-loss rate with the Eddington factor is steeper in the \citet{graef21} recipe. When comparing our 250~\msun\ track with the 250~\msun\ track of \citet{chen15} at Z=0.006\footnote{This value of Z is the closest in the Chen et al. grid to our adopted value, Z=0.0055.} , we find that on the main sequence, our mass-loss rates are 0.3~dex higher on average than those of the Chen et al. track.

According to \citet{senchyna21}, the updated Bruzual \& Charlot models rely on the PoWR\footnote{\url{https://www.astro.physik.uni-potsdam.de/~wrh/PoWR/powrgrid1.php}} grids of Wolf-Rayet models for synthetic spectra \citep{todt15}. Hence dedicated VMS synthetic spectra are not included. In particular, this implies that the mass-loss rates and helium mass fraction used for the synthetic spectra are not consistent with those of the evolutionary tracks of \citet{chen15}. Nonetheless, \citet{senchyna21} argued that VMSs are necessary to properly reproduce the UV spectrum of extreme star-forming galaxies. In their models, the EW of \heiiuv\ and \civuv\ can only be matched with VMSs. The absorption (emission) EW of line \civuv\ (\heiiuv) in our population synthesis models is 8.6 (-4.7)~\AA\ at 1~Myr and 8.3 (-9.0~\AA) at 2~Myr. From Fig.~4 of \citet{senchyna21}, this is even larger than what the Bruzual \& Charlot models predict, although they use a constant formation history rather than a burst of star formation, which partly explains the difference. The differences in mass-loss rates may also partly explain that we predict stronger lines.

\subsection{Young massive clusters}
\label{s_popsynclu}

In this section we now investigate whether our population synthesis models allow a better fit to the observed spectrum of young massive clusters.

\begin{figure}[t]
\centering
\includegraphics[width=0.49\textwidth]{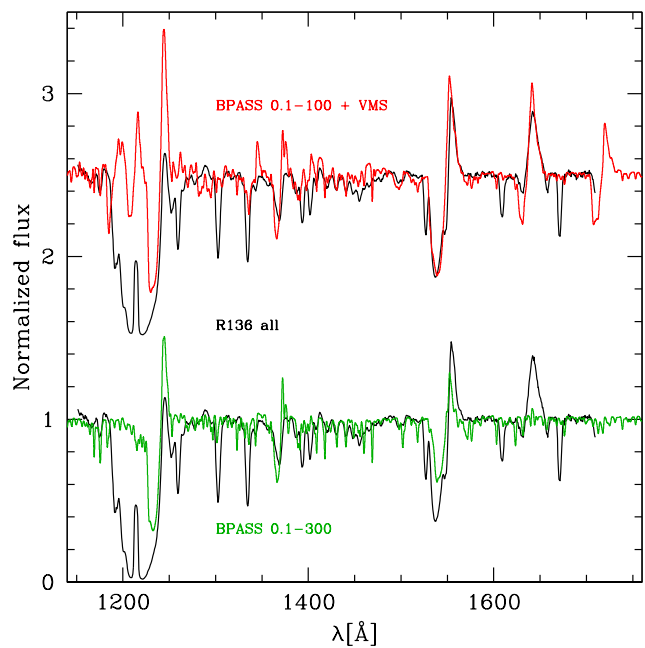}
\caption{Normalized UV spectrum of the R136 cluster (black) with our population synthesis model including VMS at 1~Myr (top, red) and the \emph{BPASS} model for an upper mass cut-off of 300~\msun\ (bottom, green). The synthetic spectra have been normalized and degraded to a spectral resolution of 1500.}
\label{r136all}
\end{figure}

The first direct test is shown in Fig.~\ref{r136all}. We compare the integrated UV spectrum of the R136 cluster in the LMC \citep{crowther16} to our simple population synthesis. We also show the \emph{BPASS} predictions with an upper mass cut-off of 300~\msun. Our test model matches the observed \heiiuv\ better than the \emph{BPASS} spectrum. \civuv\ is also stronger in our model and agrees better with the observed profile. \ion{O}{iv}~1340 and \ion{O}{v}~1371 are slightly too strong in our model compared to the spectrum of R136. The comparison of \ion{Si}{iv}~1400 is not straightforward without an assessment of the interstellar absorption. We note that the IMF of the 30~Dor region, of which R136 is part, may be top heavy \citep{schneider18}, while we used a Salpeter mass function. Changing the IMF would increase the contribution of VMS and strengthen some of the emission lines. In spite of these shortcomings, we conclude that the morphology of the spectrum including VMS is closer overall to what is observed than the pure \emph{BPASS} spectrum.

\begin{figure}[t]
\centering
\includegraphics[width=0.49\textwidth]{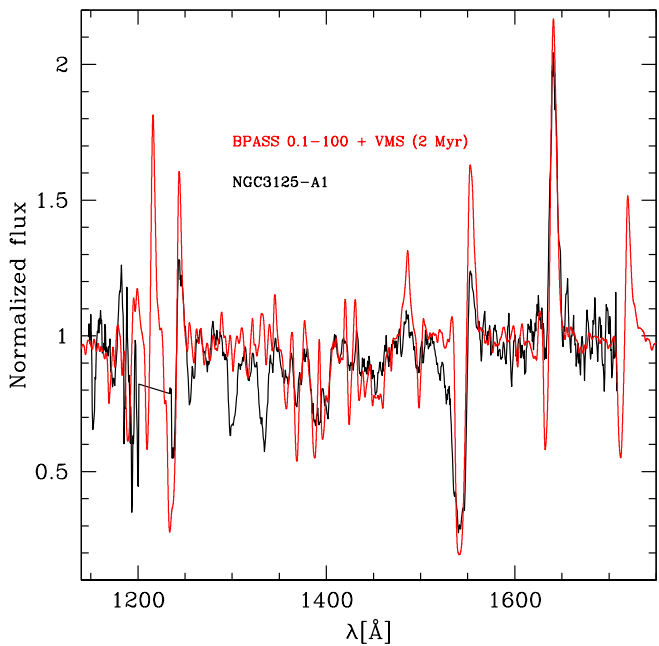}
\caption{Normalized UV spectrum of the NGC3125~A1 cluster (black) with our population synthesis model including VMS at 2~Myr (red). The synthetic spectrum has been normalized and degraded to a spectral resolution of 1000 typical of the STIS/G140L grating.}
\label{ngc3125}
\end{figure}

We further investigated whether our improved population synthesis model might reproduce the integrated spectrum of the super star cluster A1 in NGC3125 better. \citet{wofford14} showed that some of the strongest UV features, including \heiiuv, could not be reproduced by classical \emph{STARBURST99} models \citep{leitherer14}. A comparison with our models is relevant since the metallicity of NGC3125 is estimated to be close to that of the LMC \citep{hadfield06}. We retrieved the HST/STIS G140L spectrum of the cluster from the MAST archive\footnote{\url{https://archive.stsci.edu/hst/search.php}}. The spectrum has a resolution of about 1000 and covers the range 1150-1700~\AA. We normalized the spectrum and cut the spectral region around \lya\ because it is strongly absorbed. We show how our population synthesis models including VMS at 2~Myr compares to the spectrum of NGC3125~A1 in Fig.~\ref{ngc3125}. The agreement is good, with \ion{Si}{iv}~1400 and \heiiuv\ being relatively well accounted for given that we do not explore the full parameter space (age, IMF shape, star formation history). In particular, we are able to reproduce the broad \heiiuv\ emission, which was the main problem encountered by \citet{wofford14}.
\ion{N}{v}~1240 and \ion{N}{iv}~1486 are present in both our model and the observation, but we predict a slightly too strong emission. As already discussed, this may be due to the adopted initial nitrogen content of our models. Unfortunately, \nivuv\ is not observed and cannot be used to investigate this issue. We also note that \ion{N}{iv}~1486 is very sensitive to age, appearing and increasing after $\sim$1.5~Myr (see Fig.~\ref{sv200}). \civuv\ is also slightly too strong in our model. Future investigation with proper population synthesis models will help understand if this can be solved by tuning model parameters. The key points of our comparison are that we can reproduce \heiiuv\ without invoking an extremely top-heavy mass function, as in \citet{wofford14}, and that we predict \ion{N}{iv}~1486 emission (although it is too strong). 

\citet{hadfield06} studied the optical spectrum of the clusters in NGC3125. They used the classical blue and red bumps near 4650 and 5800~\AA,\ respectively, to infer the ratio of WN to WC stars. They relied on template spectra of classical Wolf-Rayet stars. Here we caution that the presence of VMSs may alter the WN/WC ratios determined by \citet{hadfield06} since both bumps are present in the spectra of VMSs. VMSs could thus contribute a significant fraction of the flux at the corresponding wavelength, reducing the contribution of classical WR stars. Here again, future studies with population synthesis models including VMSs are necessary.
From our population synthesis experiments and comparisons to the spectra of young massive clusters, we conclude that it is important to include a proper treatment of VMS evolution and atmospheres to correctly predict the integrated light of young starburst, especially in the UV spectral range.

\section{Concluding remarks}
\label{s_conc}

We have presented evolutionary tracks for VMSs including the mass-loss recipe of \citet{graef21}. We used the code \emph{STAREVOL}. We focused on the main sequence for initial masses of 150, 200, 250, 300, and 400~\msun. These tracks were compared to previous calculations and agree well. VMSs loose a significant amount of mass during their main-sequence evolution, so that they exhibit CNO-cycle processed material, including large amounts of helium, very early on during this evolutionary phase.

We subsequently computed atmosphere models with the code \emph{CMFGEN}. We selected points equally spaced in age along each evolutionary track. We showed that VMSs have an optical and UV spectrum that is dominated by \ion{He}{ii} lines for almost the entire main sequence. \heiiopt\ is the strongest optical line in all phases, and \heiiuv\ is stronger than any other UV line shortly after the star leaves the ZAMS. In the optical range, nitrogen lines are the second characteristic features. Balmer hydrogen lines are predicted in emission for most of the evolution, making the spectra of our VMS models typical of WNh stars. In the UV, a well-developed \nivuv\ P-Cygni profile, the presence of broad features caused by the \ion{Fe}{v} forest, and emission in \ion{N}{iv}~1486 are features specific to VMSs. Our synthetic spectra are morphologically similar to stars RMC~146 and R136~a1 in the LMC. 

We tested the impact of VMS on the integrated spectrum of a young starburst. We added the contribution of VMSs with masses between 100 and 300~\msun\ to a \emph{BPASS} population synthesis model at Z=0.006. We showed that taking the spectra of VMSs properly into account is crucial, especially in the UV range, where VMSs dominate the integrated spectrum: \heiiuv\ emission is entirely caused by the presence of VMSs, \nivuv\ is much stronger when VMSs are included; \civuv\ is also affected; and VMSs produce \ion{N}{iv}~1486 emission after $\sim$1-1.5~Myr. VMSs contribute almost as much light as the rest of the population in that wavelength range. Our population synthesis test model including VMSs reproduces the integrated spectrum of R136 better than the \emph{BPASS} model that extends to 300~\msun. The reason is the proper treatment of mass loss in both the evolutionary models and synthetic spectra of VMSs. Including VMSs also helps to reproduce the strong UV lines of the super star cluster A1 in NGC3125.

In our study we have worked with single stars. Some of the most massive stars known to date are members of binary systems:  NGC3602-A1 \citep{schnurr08}, R145 \citep{schnurr09}, Arches-F2 \citep{lohr18}, Melnick~34 \citep{tehrani19}, and R144 \citep{shenar21}. At least one of the components usually shows a spectrum typical of a WNh star. Including binary stars in population synthesis models extending up to 300~\msun\ is thus necessary. However, we anticipate that our conclusion regarding the need for a proper treatment of the wind physics of VMSs remains at least qualitatively valid. Integrated spectra of young star-forming galaxies are usually obtained at relatively low spectral resolution where the component of the massive binary systems listed above are blended. The light is thus dominated by the most massive component, or in case of nearly equal-mass systems, is the sum of the two VMS components. 

In this study we have focused on a metallicity Z~=~1/2.5~\zsun\ that can be viewed as typical of the LMC. We chose this metallicity because the mass-loss recipe for VMSs was mainly established from the analysis of LMC stars \citep{besten20,graef21}. The LMC also hosts the most massive stars that can be observed individually and to which we can compare our predictions. In future studies, we will extend our predictions to a wider range of metallicities. We will also test the ability of our population synthesis models to reproduce the UV spectrum of young star-forming galaxies, in particular, their stellar \heiiuv\ emission \citep[e.g.][]{erb10,saxena20a,marques20,marques21}.

\section*{Acknowledgments}

We thank an anonymous referee for a positive report and suggestions of clarification. We warmly thank John Hillier for making the code \emph{CMFGEN} available to the community and for constant help with it. This work made use of v2.2.1 of the Binary Population and Spectral Synthesis (\emph{BPASS}) models as last described in \cite{staneldr18}. We thank Paul Crowther and Joachim Bestenlehner for sharing the UV spectra of R136~a1 and of the entire R136 population. We acknowledge interesting discussions with Daniel Schaerer and Rui Marques-Chaves about the UV spectra of young starburst galaxies. Some of the data presented in this paper were based on observations made with the NASA/ESA HST, obtained from the MAST data archive at the Space Telescope Science Institute (STScI). The STScI is operated by the association of Universities for Research in Astronomy, Inc.
under the NASA contract NAS 5-26555. These observations are associated with program 12172. This research has made use of NASA’s Astrophysics Data System.

\bibliographystyle{aa}
\bibliography{vms_1}

\newpage

\begin{appendix}
\label{ap_tab}

\section{Synthetic spectra throughout all mass sequences}
\label{ap_spec}

\begin{figure*}[h]
\centering
\includegraphics[width=0.49\textwidth]{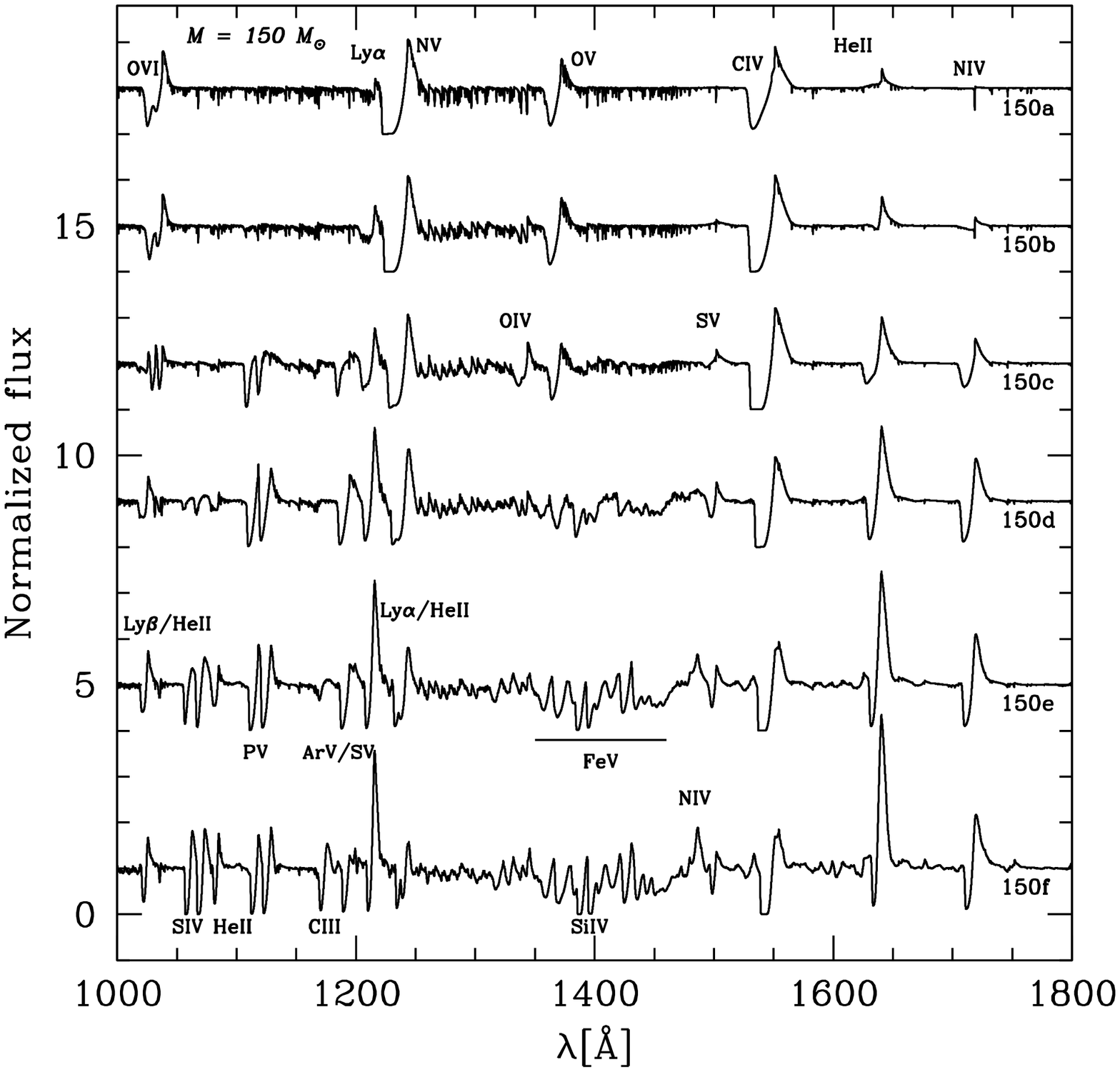}
\includegraphics[width=0.49\textwidth]{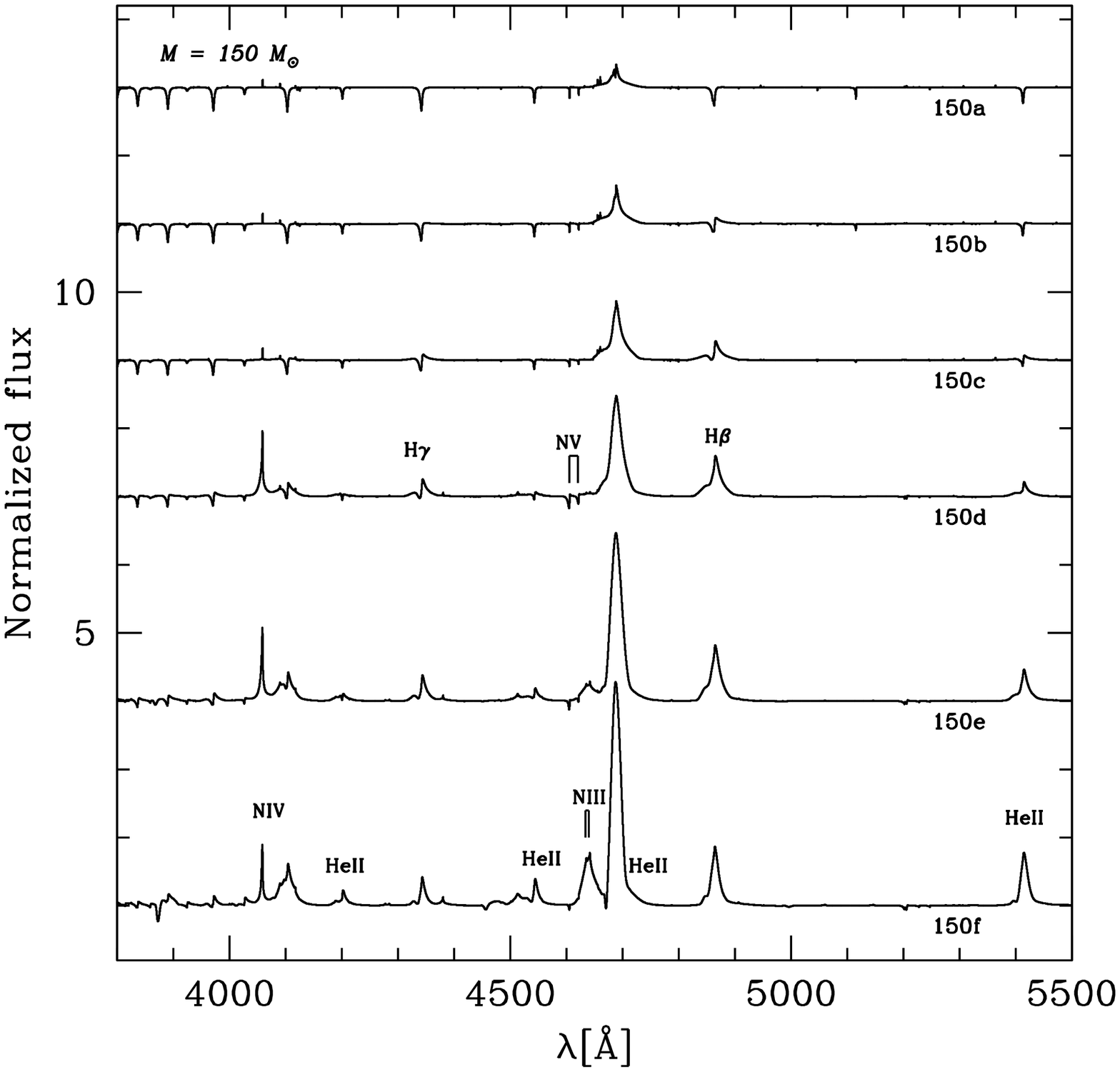}
\caption{Synthetic UV (left) and optical (right) spectra of the 150~\msun\ models.}
\label{sv150}
\end{figure*}

\begin{figure*}[h]
\centering
\includegraphics[width=0.49\textwidth]{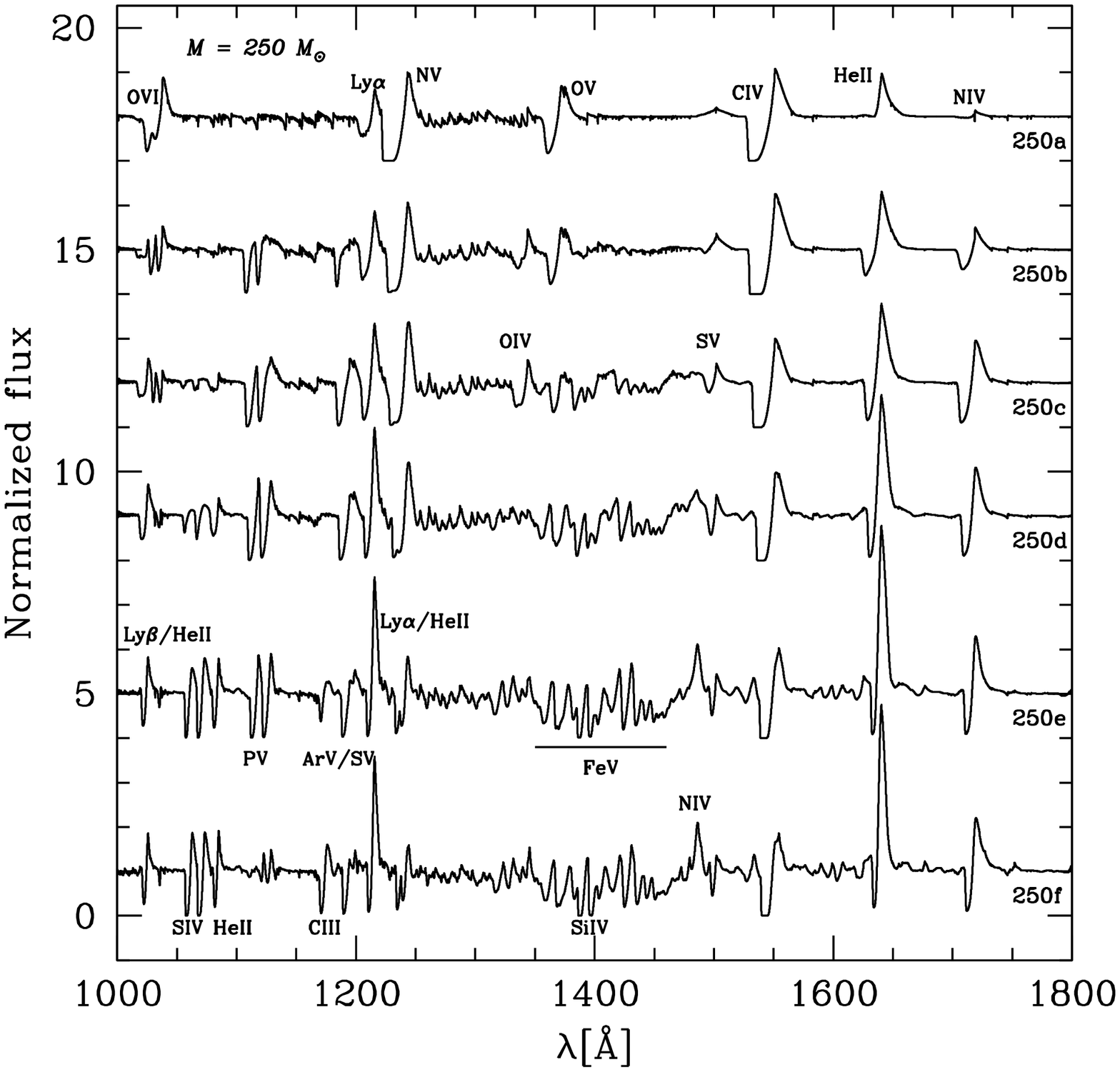}
\includegraphics[width=0.49\textwidth]{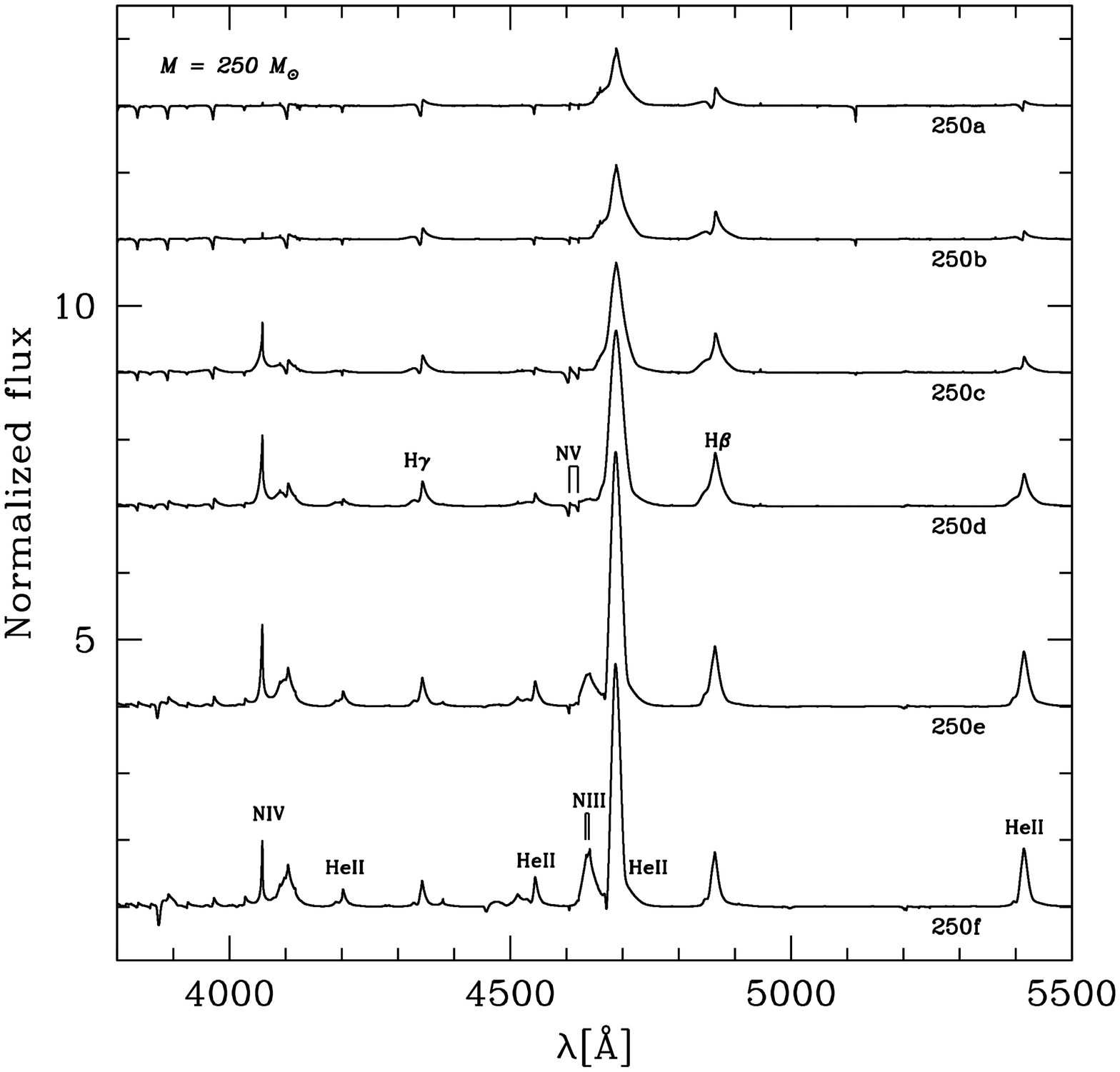}
\caption{Synthetic UV (left) and optical (right) spectra of the 250~\msun\ models.}
\label{sv250}
\end{figure*}

\begin{figure*}[h]
\centering
\includegraphics[width=0.49\textwidth]{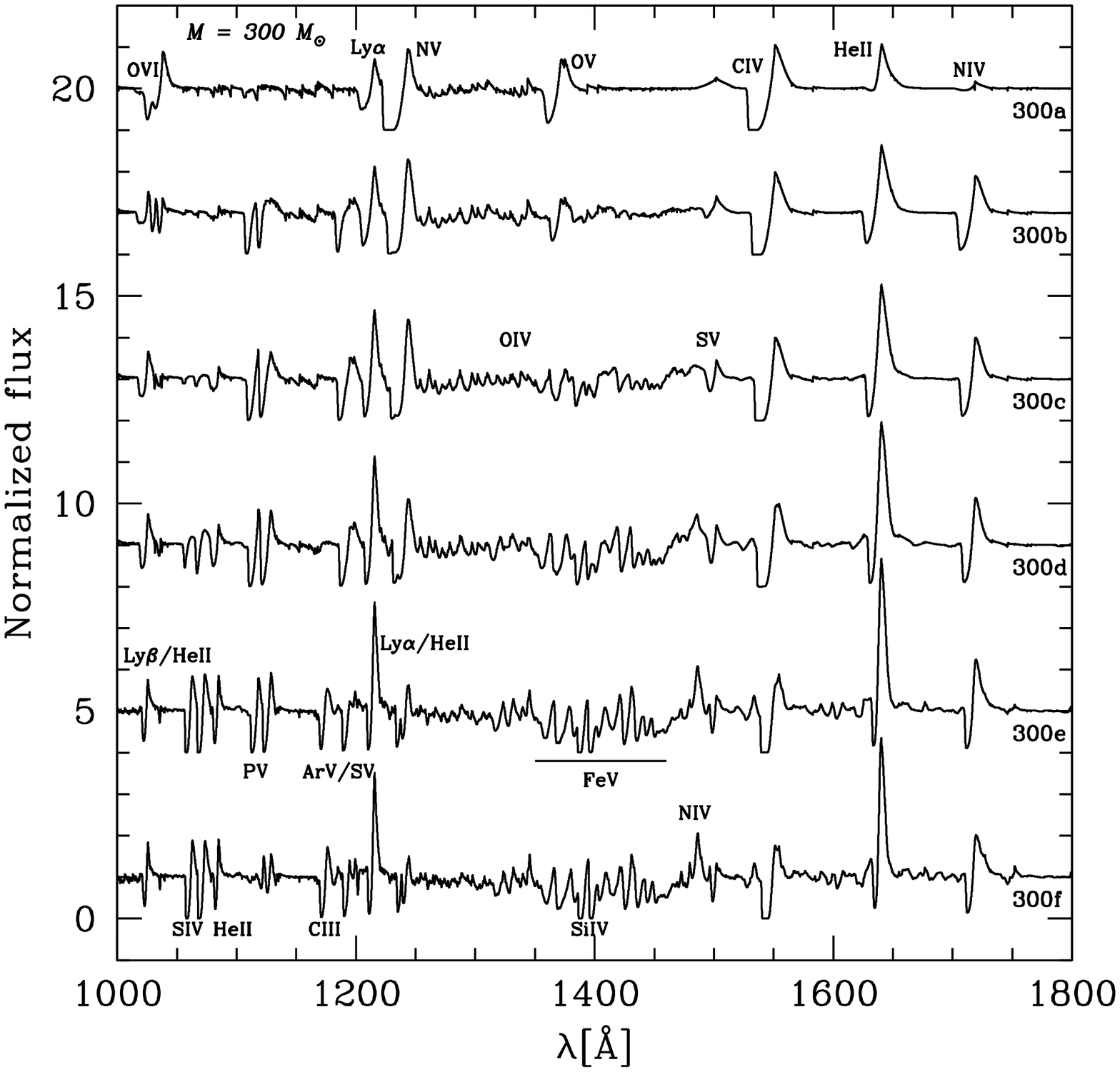}
\includegraphics[width=0.49\textwidth]{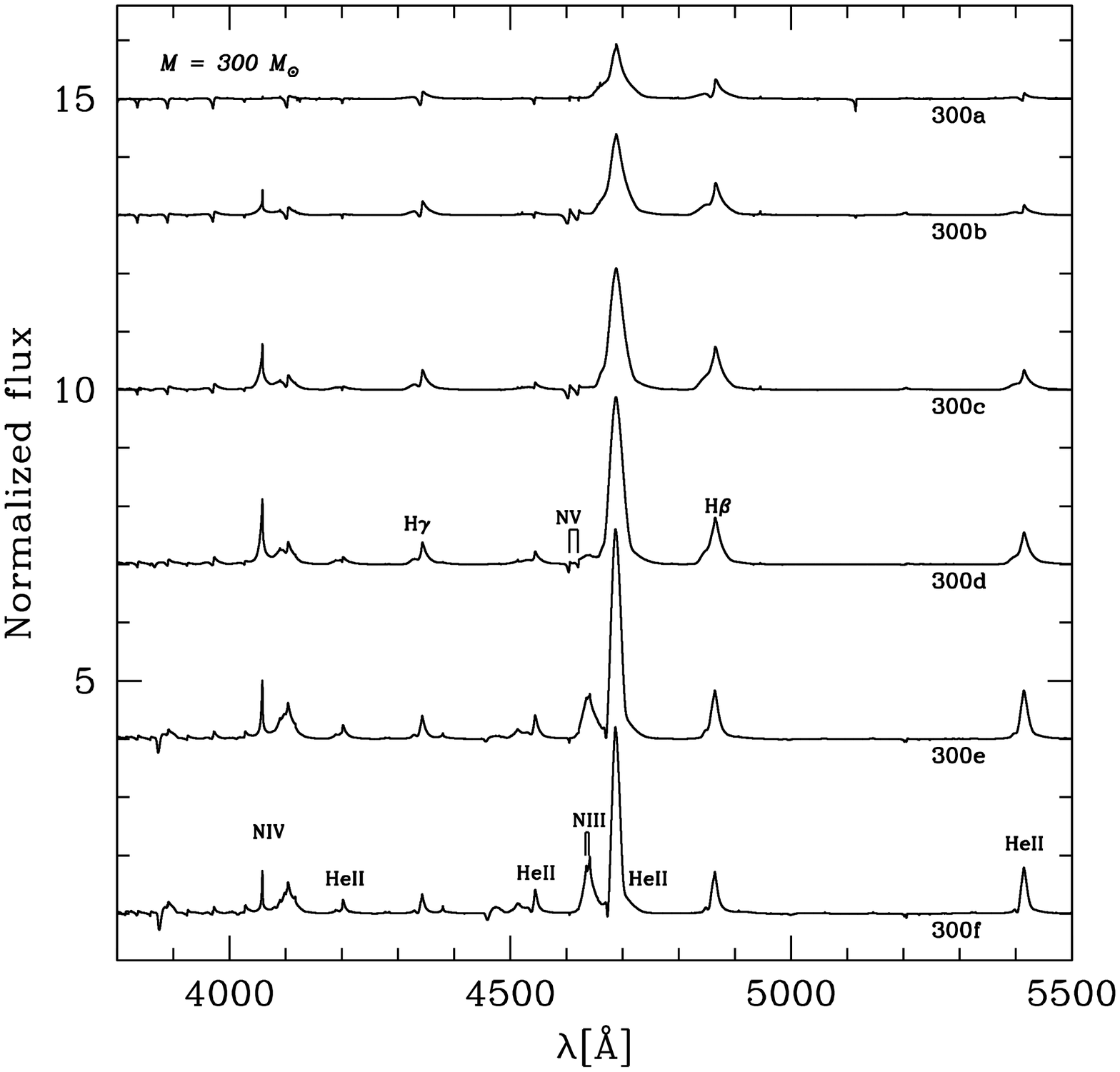}
\caption{Synthetic UV (left) and optical (right) spectra of the 300~\msun\ models.}
\label{sv300}
\end{figure*}

\begin{figure*}[h]
\centering
\includegraphics[width=0.49\textwidth]{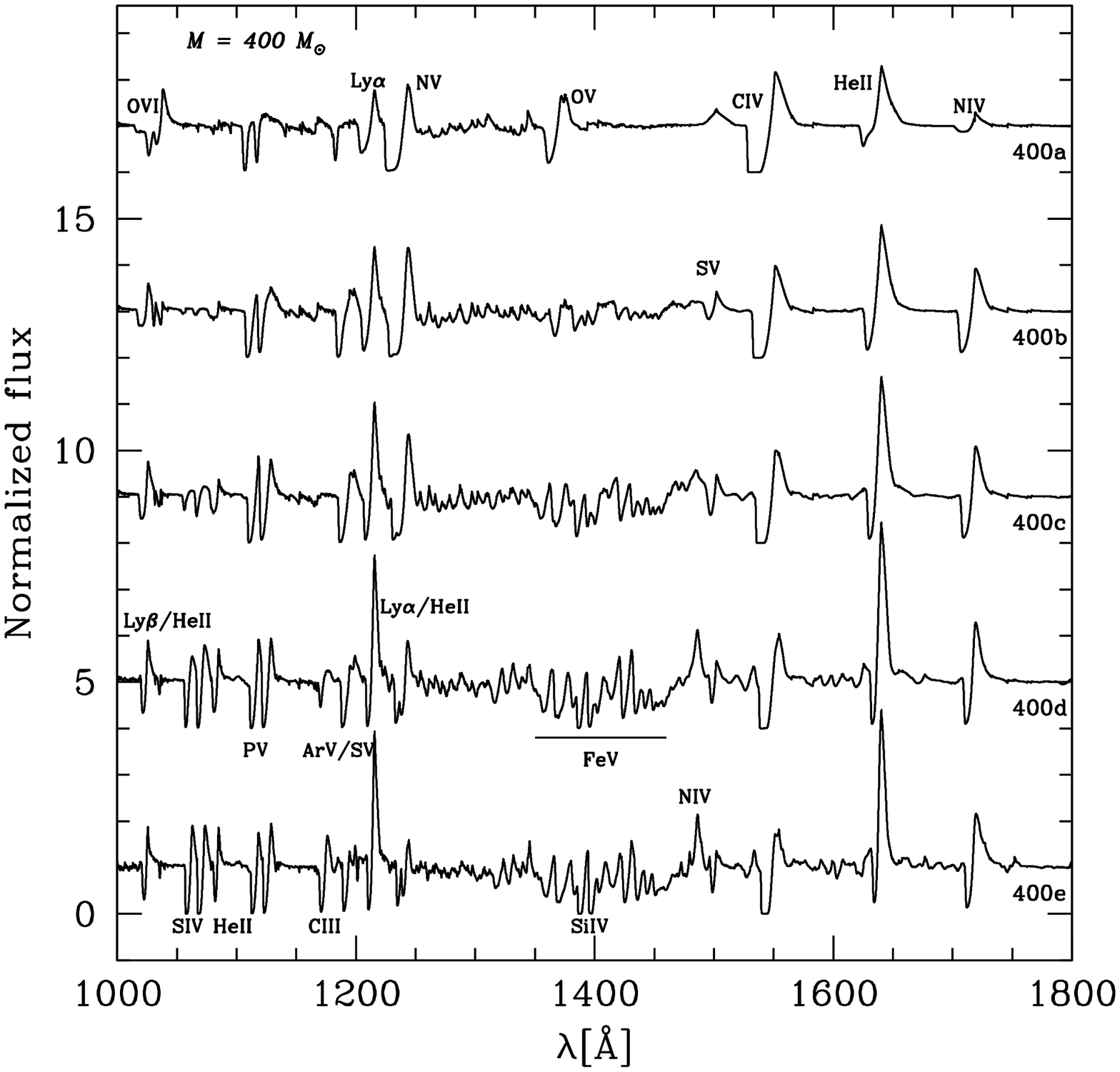}
\includegraphics[width=0.49\textwidth]{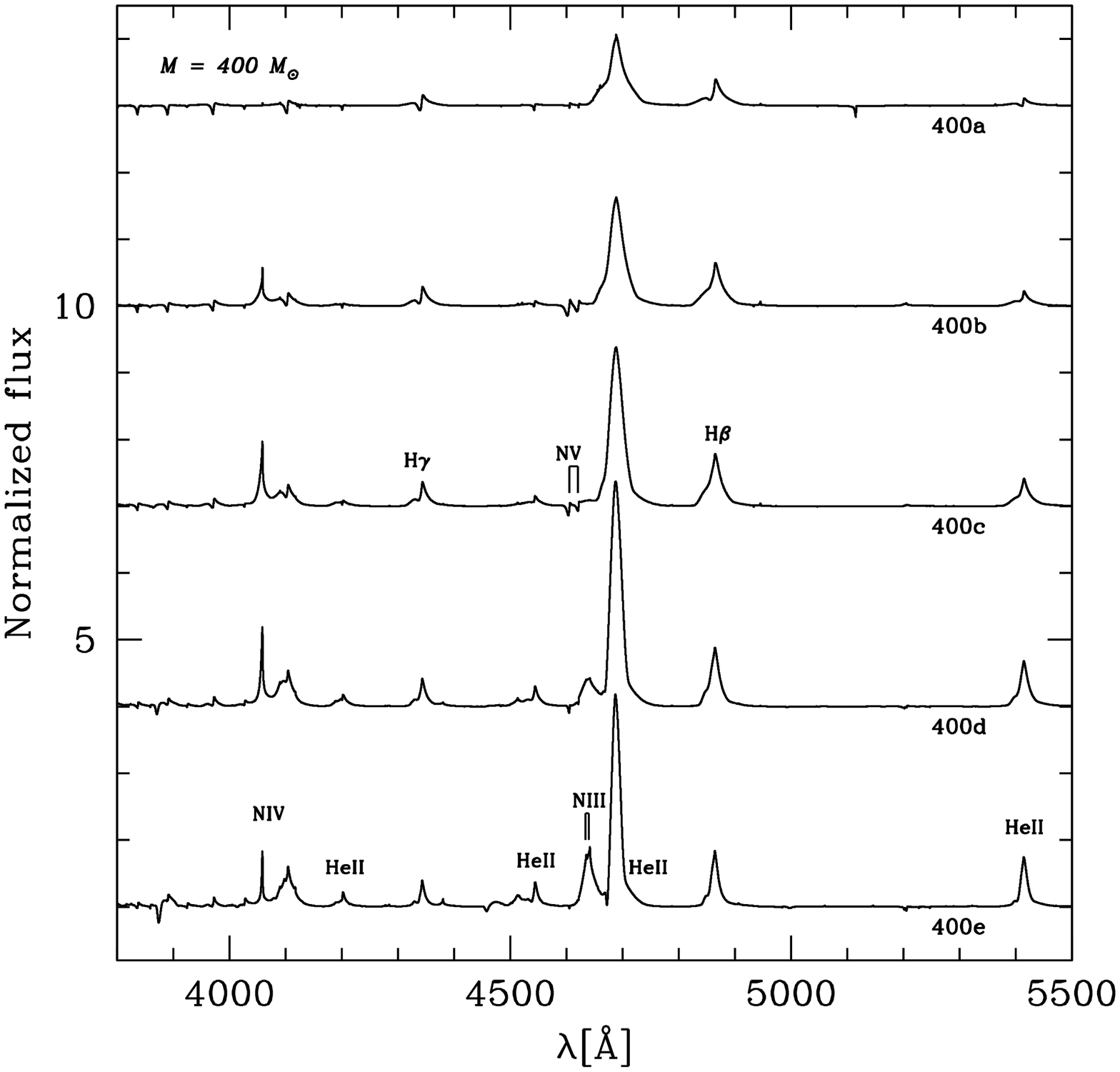}
\caption{Synthetic UV (left) and optical (right) spectra of the 400~\msun\ models.}
\label{sv400}
\end{figure*}

In Figs.~\ref{sv150} to \ref{sv400} we show the UV and optical spectra along the 150, 250, 300, and 400~\msun\ evolutionary tracks.

\FloatBarrier

\newpage

\section{Population synthesis}
\label{ap_imf}

The \emph{BPASS} SEDs are given for a cluster containing $10^6$~\msun. According to the \emph{BPASS} user manual\footnote{\url{https://flexiblelearning.auckland.ac.nz/bpass/9.html}} , the number of stars between 0.1~\msun\ and a maximum mass M$_{2}$ is

\begin{equation}
N(0.1,M_{2}) = C \times\ \left( \int_{0.1}^{M_{1}} M^{\alpha_{1}} \,dM + M_{1}^{\alpha_{1}} \int_{M_{1}}^{M_{2}} M^{\alpha_{2}} \,dM \right)
.\end{equation}

\noindent For our comparisons, we selected the \emph{BPASS} models with M$_{1}$~=~0.5~\msun, $\alpha_{1}$~=~-1.3, $\alpha_{2}$~=~-2.35, and M$_{2}$~=~100~\msun. 
The total mass of the cluster is given by 

\begin{equation}
M(0.1,M_{2}) = C \times\ \left( \int_{0.1}^{M_{1}} M^{1+\alpha_{1}} \,dM + M_{1}^{\alpha_{1}} \int_{M_{1}}^{M_{2}} M^{1+\alpha_{2}} \,dM \right)
.\end{equation}

\noindent The normalization constant C has a value of 1.23$\times$10$^5$ for a total mass equal to 10$^6$. For this set up, the number of stars with masses above 100~\msun\ and in the mass bin [M$_{a}$;M$_{b}$] is 

\begin{equation}
N(M_{a},M_{b}) = C \times\ M_{1}^{\alpha_{1}} \int_{M_{a}}^{M_{b}} M^{\alpha_{2}} \,dM 
,\end{equation}

\noindent which gives 236.56, 60.30, 35.42, and 12.16 stars for the mass bins 100-175, 175-225, 225-275, and 275-300~\msun\ , respectively. To take VMSs (M$>$100~\msun) in population synthesis into account, we thus added the spectra of the 150, 200, 250, and 300~\msun\ multiplied by these numbers to the SED provided by \emph{BPASS} (for a cluster hosting stars with masses between 0.1 and 100~\msun). We did this at two ages: 1 and 2~Myr.

\end{appendix}

\end{document}